\documentclass[letterpaper, 10 pt, conference]{ieeeconf}  

\IEEEoverridecommandlockouts                              

\usepackage{graphics} 
\usepackage{epsfig} 
\usepackage{mathptmx} 
\usepackage{times} 
\usepackage{amsmath} 
\usepackage{amssymb}  

\usepackage[english]{babel}
\usepackage[T1]{fontenc}
\usepackage[utf8]{inputenc}

\usepackage{amsfonts,amsmath,bm,hhline}
\usepackage{epsfig}
\usepackage{color}
\usepackage[hang]{caption2}
\usepackage[normalsize,it,hang]{subfigure}
\newtheorem{remark}{Remark}

\DeclareGraphicsExtensions{.jpg,.png,.jpg,.jpg}

\usepackage{algorithm}
\usepackage{algpseudocode}
\usepackage{siunitx}
\vspace{2cm}
\title{{\small \tt IEEE 2025 - 13th International Conference on Systems and Control (ICSC) \\ October 22-24, 2025 - Marrakesh, Morocco} \\ \vspace{2cm}
{\LARGE \bf Avoidance of an unexpected obstacle without reinforcement learning: \\ Why not using advanced control-theoretic tools?}}

\author{C\'edric Join$^{\dag,\S}$ and Michel Fliess$^{\P,\ddag\,\S}$
\thanks{$^{\dag}$CRAN (CNRS, UMR 7039)), Universit\'{e} de Lorraine, BP 239, 54506 Vand{\oe}uvre-l\`{e}s-Nancy, France (e-mail: \texttt{cedric.join@univ-lorraine.fr}).}%
\thanks{$^{\P}$ LIX (CNRS, UMR 7161), \'Ecole polytechnique, 91128 Palaiseau, France (e-mail: \texttt{michel.fliess@polytechnique.edu}).}
\thanks{$^{\ddag}$ LJLL (CNRS, UMR 7598), Sorbonne Universit\'{e}, 75005 Paris, France (e-mail: \texttt{michel.fliess@sorbonne-universite.fr}).}
\thanks{$^{\S}$ AL.I.E.N., 7 rue Maurice Barr\`{e}s, 54330 V\'{e}zelise, France (e-mail: \{\texttt{cedric.join, michel.fliess}\}\texttt{@alien-sas.com}).} 
}

\begin{document}
\onecolumn
\thispagestyle{empty}
\pagestyle{empty}
\maketitle

\begin{abstract}
This communication on collision avoidance with unexpected obstacles is motivated by some critical appraisals on reinforcement learning (RL) which ``requires ridiculously large numbers of trials to learn any new task'' (Yann LeCun). We use the classic Dubins' car in order to replace RL with flatness-based control, combined with the HEOL feedback setting, and the latest model-free predictive control approach. The two approaches lead to convincing computer experiments where the results with the model-based one are only slightly better. They exhibit a satisfactory robustness with respect to randomly generated mismatches/disturbances, which become excellent in the model-free case. Those properties would have been perhaps difficult to obtain with today's popular machine learning techniques in AI. Finally, we should emphasize that our two methods require a low computational burden.

\end{abstract}
\begin{keywords}
Robotics, collision avoidance, unexpected obstacle, flatness-based control, HEOL, model-free predictive control. 
\end{keywords}

\newpage 
\section{Introduction}
\emph{Reinforcement learning} (\emph{RL}) is a key technique in machine learning (see, e.g., \cite{rl}). It plays an important r\^{o}le in robotics (see, e.g., \cite{kober,zhang,drl}). A striking illustration is provided by a recent publication \cite{deepmind} from Google DeepMind on soccer skills for a bipedal robot. There are a number of ways to implement RL, and each one has its own advantages and disadvantages (see, e.g., \cite{grizzle}). On the other hand, LeCun \cite{lecun1} writes: {\it RL requires ridiculously large numbers of trials to learn any new task}. This view cannot be denied: See also \cite{lecun2,dulac}. The aim of this paper is to move beyond the overwhelming preeminence of optimization techniques in RL (see, e.g., \cite{bertsekas,recht,kober,zhang}), which are over 60 years old, by investigating new control-theoretic tools that have perhaps not been previously sufficiently explored:
\begin{enumerate}
    \item (\emph{Differentially}) \emph{flat} systems \cite{flmr1,flmr2}, which are popular in many concrete domains, but unfortunately unknown in the world of AI even when connected to robotics, and the new corresponding {\em HEOL} feedback setting \cite{heol}, which has been recently illustrated \cite{wien,degorre,mfpc}.
    \item {\em Model-free predictive control} ({\em MFPC}), as defined in \cite{mfpc}, already provides some connections with AI. It is based on
    \begin{itemize}
        \item recent advances \cite{mfc1,mfc2} in \emph{model-free control} (\emph{MFC}) which found many applications (see, e.g., among recent contributions \cite{fenyes,liu,ait,acm}, and \cite{oak} for collision avoidance without reinforcement learning);
        \item a new understanding \cite{batna,cocv} of optimal control without tedious calculations. 
    \end{itemize}
\end{enumerate}
Our approach is illustrated by an example borrowed from \cite{norv}, where RL is the main ingredient. A simple autonomous wheeled robot, which is often encountered under the name of \emph{Dubins' car} (see, e.g., \cite{dubin}), has to avoid collisions with unexpected obstacles. This question is today a major issue in robotics. Let us stress the following points on our computer simulations:
\begin{itemize}
    \item The performance with the model-based approach (flatness-based control $+$ HEOL) is, of course, better than with MFPC, although the last ones are more than acceptable.
    \item The robustness with respect to randomly generated disturbances/mismatches is perhaps better with MFPC than with the combination of flatness-based setting and HEOL. How would popular machine learning techniques in AI behave in such a situation?
    \item Easy implementation and low computational burden.
\end{itemize}
Our paper is organized as follows. After introducing our autonomous robot, Sect. \ref{flat} shows its flatness and, thanks to a change of variables, presents a most elementary HEOL feedback setting. The basics of MFPC are outlined in Sect. \ref{gif}. Sect. \ref{traj} develops the main aspect of this study, namely the obstacle avoidance. The diagram of Fig. \ref{algo} should hopefully convince the readers of the efficiency and simplicity of our viewpoint. See Sect. \ref{conclu} for some concluding remarks.

\section{The flatness-based approach}\label{flat}
\subsection{Model description}
\begin{figure}[!ht]
{\epsfig{figure=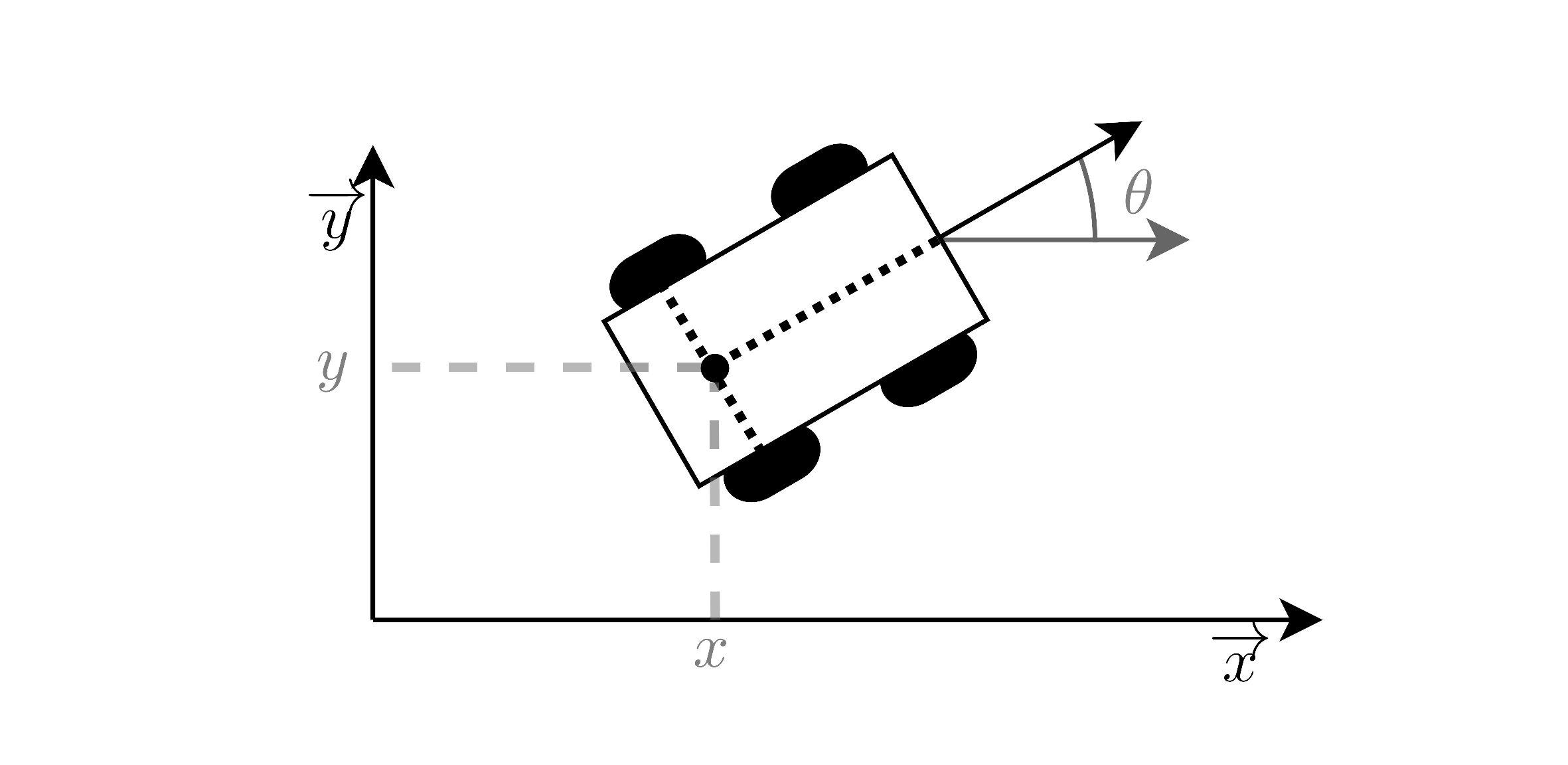,width=0.9\textwidth}}
\centering
\caption{Our autonomous robot: Dubins'car}\label{scheme}
\end{figure}

Fig. \ref{scheme} shows an autonomous wheeled robot with two degrees of freedom. Its dynamics is given by
\begin{equation}\label{MV}
\begin{cases}
    \dot x=u_1\cos(u_2)\\
    \dot y=u_1\sin(u_2)
\end{cases}
\end{equation}
where 
\begin{itemize}
    \item $(x, y)$ are the Cartesian coordinates of the middle of the rear axle; 
    \item the control variables $u_1$ and $\theta = u_2$ are respectively the linear velocity, and the angle between the heading direction and the $x$ axis. 
\end{itemize}
\subsection{Flatness}
Follow \cite{degorre} and introduce the auxiliary control variables
$\nu_1=u_1\cos(u_2)$ and $\nu_2=u_1\sin(u_2)$. It yields for Eq. \eqref{MV} two parallel pure integrators: 
\begin{equation}\label{paral}
\begin{cases}
    \dot x=\nu_1\\
    \dot y=\nu_2
\end{cases}
\end{equation}
It shows at once that our robot is flat and that $x$, $y$ are flat outputs (see also \cite{flmr1}). 
\subsection{HEOL}
The HEOL setting \cite{heol} becomes trivial. Eq. \eqref{paral} yields the {\em homeostat} \cite{heol}: 
\begin{equation}\label{homeo}
\begin{cases}
    \frac{d}{dt} (\Delta x) = F_x + \Delta \nu_1\\
    \frac{d}{dt} (\Delta y) = F_y + \Delta \nu_2
\end{cases}
\end{equation}
where 
\begin{itemize}
    \item $\Delta \zeta = \zeta - \zeta_{\rm ref}$, 
    \begin{itemize}
        \item $\zeta$ stands for $x$, $y$, $\nu_1$, or $\nu_2$, 
        \item $\zeta_{\rm ref}$ is the open-loop variable corresponding to a reference trajectory,
    \end{itemize}
    \item $F_x$ and $F_y$ stand for the mismatches and disturbances.
\end{itemize}
Thanks to Eq. \eqref{homeo}, the loop is closed via the \emph{intelligent proportional controller} (\emph{iP})
\begin{equation}\label{ip}
\begin{cases}
    \Delta \nu_1 = - (\hat{F}_x + K_{x, P} \Delta x) \\
    \Delta \nu_2 = - (\hat{F}_y + K_{y, P} \Delta y)
\end{cases}
\end{equation}
where {\scriptsize$ \hat{F}_x =  
 - \frac{6}{T^3} \int_{0}^{T} \left( (T - 2 \sigma)\Delta x(\sigma + T -\tau) {+} \sigma (T - \sigma)\Delta u_1(\sigma + T - \tau)\right)d\sigma
$} and  {\scriptsize$\hat{F}_y =  
 - \frac{6}{T^3} \int_{0}^{T} \left( (T - 2 \sigma)\Delta x(\sigma + T -\tau) {+} \sigma (T - \sigma)\Delta u_1(\sigma + T - \tau)\right)d\sigma$} 
 are the estimates of $F_x$ and $F_y$ (see \cite{heol}). The gains $K_{x, P}$ and $K_{y, P}$ are constant and positive: they insure local stability if the estimates are ``good,'' i.e., $F_x \approx \hat{F}_x$, $F_y \approx \hat{F}_y$.

Note that the true control variables are given by $u_1= \sqrt{\nu_1^2+\nu_2^2}$ and $u_2=\text{atan2}(\nu_2,\nu_1)$.\footnote{See, e.g., Wikipedia for the definition of the function atan2.}









\section{Model-free predictive control}\label{gif}
 \subsection{Generalities\protect\footnote{See \cite{mfpc}.}}
 \subsubsection{Model-free control}\label{heol}
Consider with \cite{mfc1} the SISO \emph{ultra-local model} of order $1$, which is replacing the poorly known plant and disturbances: 
\begin{equation}
\dot{y} = \mathcal{F} + \alpha u
\label{1}
\end{equation}
The derivation order of $y$ is $1$ like in many real-life situations.
$\mathcal{F}$ subsumes not only the unknown structure but also
any external disturbance. 
The constant $\alpha \in \mathbb{R}$ is chosen by the practitioner such that $\alpha u$ and $\dot{y}$ are of the same magnitude. Therefore $\alpha$ does not need to be precisely estimated. A data-driven estimation of $\mathcal{F}$ in Eq. \eqref{1} reads:{\footnotesize
\begin{equation*}\label{integral}
\begin{aligned}
 \mathcal{F}_{\text{est}}(t)  =-\frac{6}{T^3}\int_{0}^T& \left( (T -2\sigma)y(t-T+\sigma)\right.\\&\left.+\alpha(T-\sigma)\sigma u(t-T+\sigma) \right) d\sigma 
\end{aligned}
\end{equation*}}

\subsubsection{Optimal control}
Assume that $F = a$ is a constant in Eq. \eqref{1}, which corresponds now to an elementary flat system, where $y$ is a flat output. Introduce, like in \cite{batna}, the Lagrangian, or cost function, 
\begin{equation*}\label{lag}
\mathcal{L} = (y - y_{\rm setpoint})^2 + u^2 = (y - y_{\rm setpoint})^2 + \left(\frac{\dot{y} - a}{\alpha}\right)^2
\end{equation*}
where $y_{\rm setpoint}$ denotes a given setpoint. 
For the criterion
$
    J = \int_{t_i}^{t_f} \mathcal{L}dt 
$
the Euler-Lagrange equation 
$
\frac{\partial \mathcal{L}}{\partial y} - \frac{d}{dt} \frac{\partial \mathcal{L}}{\partial \dot{y}}  = 0
$
corresponds to a non-homogeneous linear ordinary differential equation of $2$nd order $\ddot{y} - \alpha^2 (y - y_{\rm setpoint}) = 0$.
Any optimal solution  reads 
$
    y^\star (t)= y_{\rm setpoint} + c_1\exp(\alpha t) + c_2\exp(-\alpha t)$, $c_1, c_2 \in \mathbb{R}$.
It is independent of $a$; $c_1$, $c_2$ are obtained via the two-point boundary conditions $y(t_i) = y_i$, $y(t_f) = y_{\rm setpoint}$: $$c_1 =\frac{y_i\exp(-\alpha t_f) - y_{\rm setpoint}\exp(-\alpha t_f)}{\exp(\alpha t_i)\exp(-\alpha t_f) - \exp(-\alpha t_i)\exp(\alpha t_f)}$$
$$c_2 =-\frac{\exp(\alpha t_f)(y_i - y_{\rm setpoint})}{\exp(\alpha t_i)\exp(-\alpha t_f) - \exp(-\alpha t_i)\exp(\alpha t_f)}$$

\subsubsection{Implementation}\label{imple}
Consider the subdivision $0< \cdots < t_k < t_{k+1} < \cdots < t_f$. On each time lapse $[t_k,t_{k+1}]$ replace $\mathcal{F}$ in Eq.~\eqref{1} by the constant $\mathcal{F}_\text{est}(t_k)$. Start the above optimization procedure again on the time horizon $[t_{k+1},t_f]$. The criterion and the horizon may be modified if necessary.

\subsection{Application to the robot}
The ultra-local model for our MIMO robot reads (see, e.g., \cite{toulon}) 
\begin{equation}\label{u}
\begin{cases}
    \dot x=F_1+\alpha_1 u_1\\
    \dot y=F_2+\alpha_2 u_2
\end{cases}
\end{equation}
where, since we are not using $\sin (u_2)$, we impose $-\pi / 2 < u_2 < \pi /2$. 
Admittedly, this limitation complicates the control. However, the coefficient $\alpha_2$ should obviously be time-varying without it. 

Implementation of MFPC is achieved via the application of Sect. \ref{imple}.

\begin{remark}
    Replacing $u_2$ in Eq. \eqref{u} by $\sin (u_2)$ would yield comparable results to those via flatness-based control. Lack of space prevents developing this approach.
\end{remark}

\section{Computer simulations}
All simulations are performed with a sampling period of $0.01$s. They last $20$s.

\subsection{Synchronizing trajectories}\label{traj}
To avoid or limit backtracking, the reference trajectories $x_{\rm ref}$ and $y_{\rm ref}$ can be revised. These modifications are activated at startup when the vehicle's initial position is far from the trajectory and after obstacle avoidance, which is developed below. These positions are defined by the coordinates $x_{\rm sync}(t)$ and $y_{\rm sync}(t)$. A time offset, $\tau$, is realized that minimizes $\left(x_{\rm sync}(t)-x_{\rm ref}(t+\tau)\right)^2+\left(y_{\rm sync}(t)-y_{\rm ref}(t+\tau)\right)^2$.

\begin{figure*}[!ht]
\centering
\subfigure[\footnotesize $x$]
{\epsfig{figure=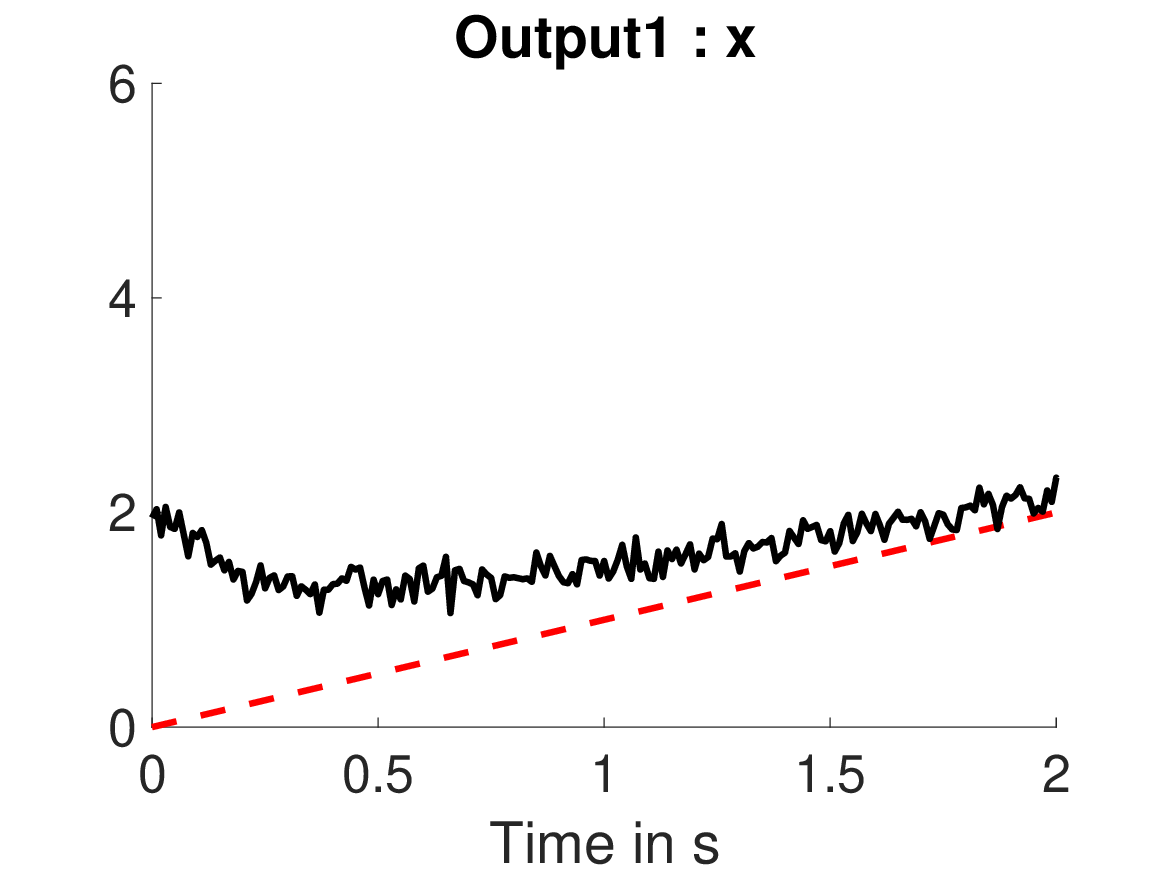,width=0.3\textwidth}}
\subfigure[\footnotesize $y$]
{\epsfig{figure=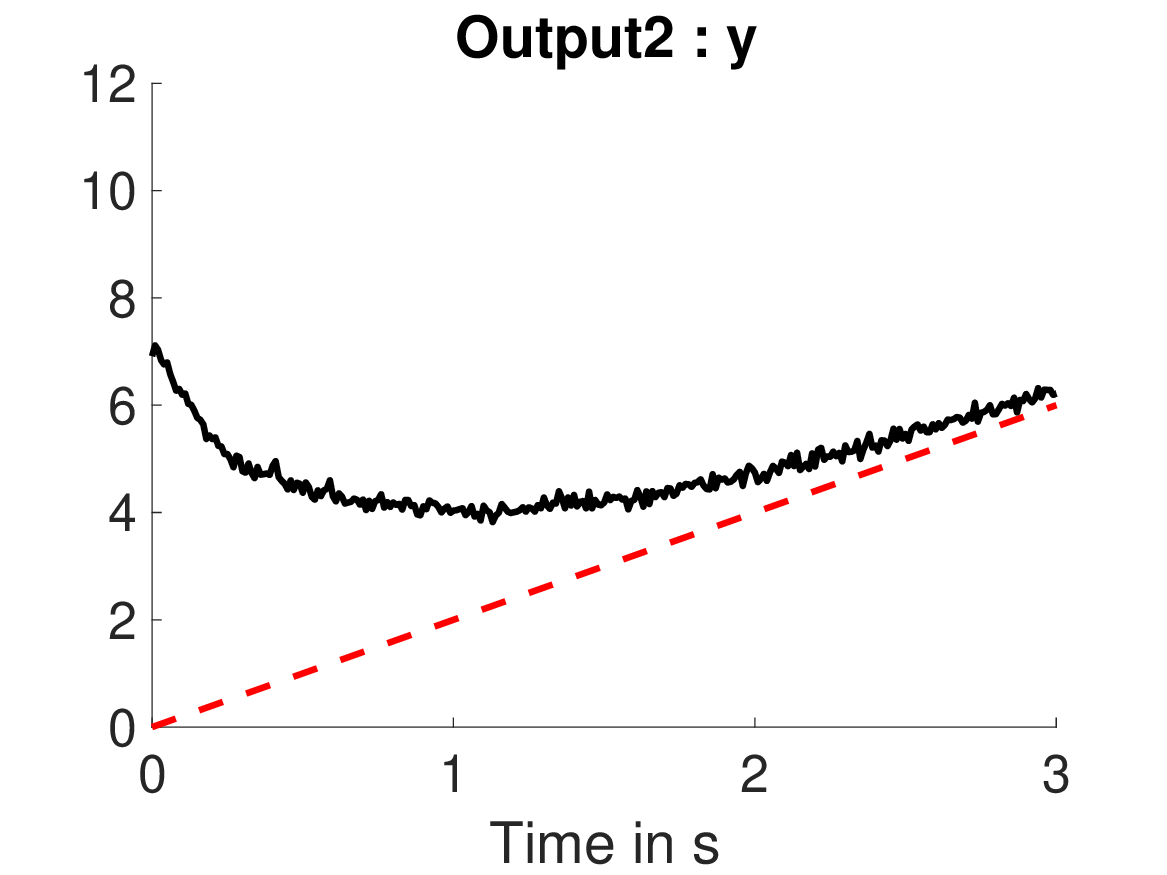,width=0.3\textwidth}}
\subfigure[\footnotesize $(x,y)$]
{\epsfig{figure=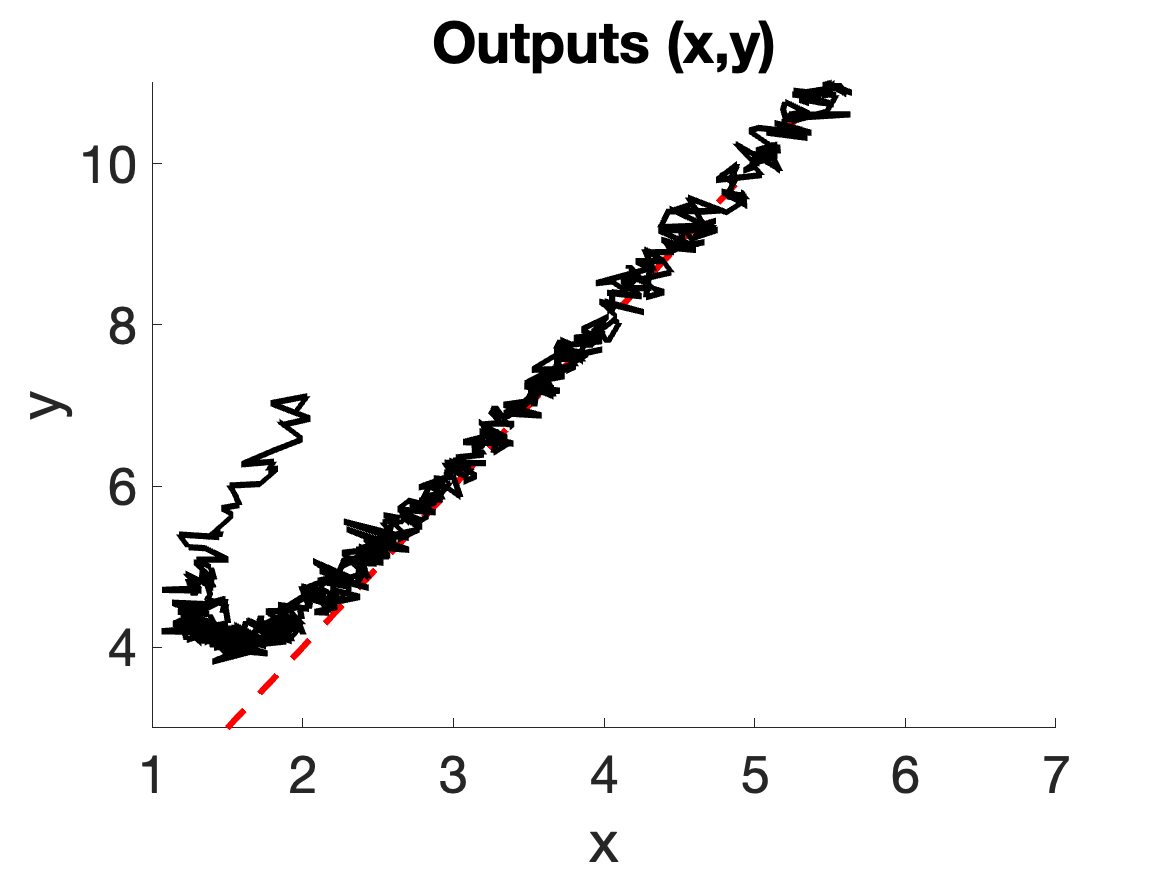,width=0.3\textwidth}}

\caption{Without trajectory synchronisation}\label{I}
\end{figure*}
\begin{figure*}[!ht]
\centering
\subfigure[\footnotesize $x$]
{\epsfig{figure=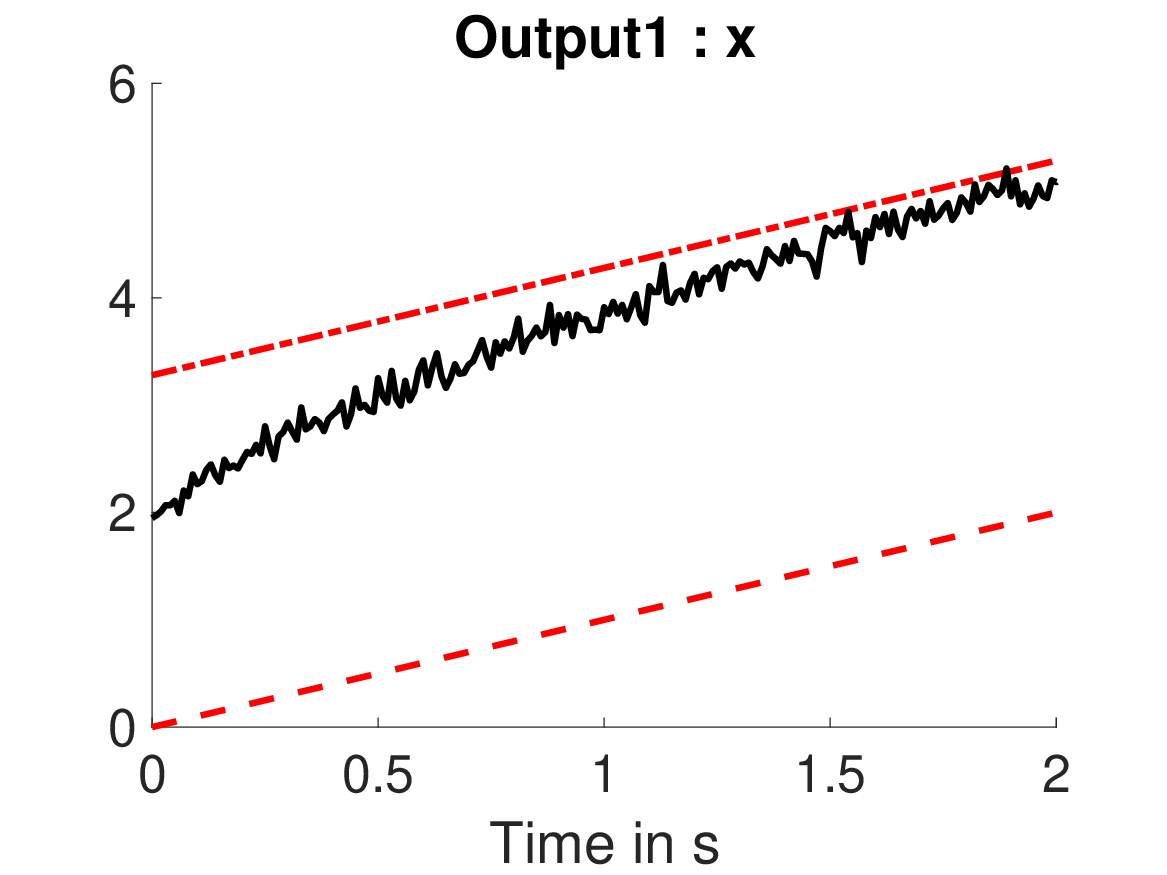,width=0.3\textwidth}}
\subfigure[\footnotesize $y$]
{\epsfig{figure=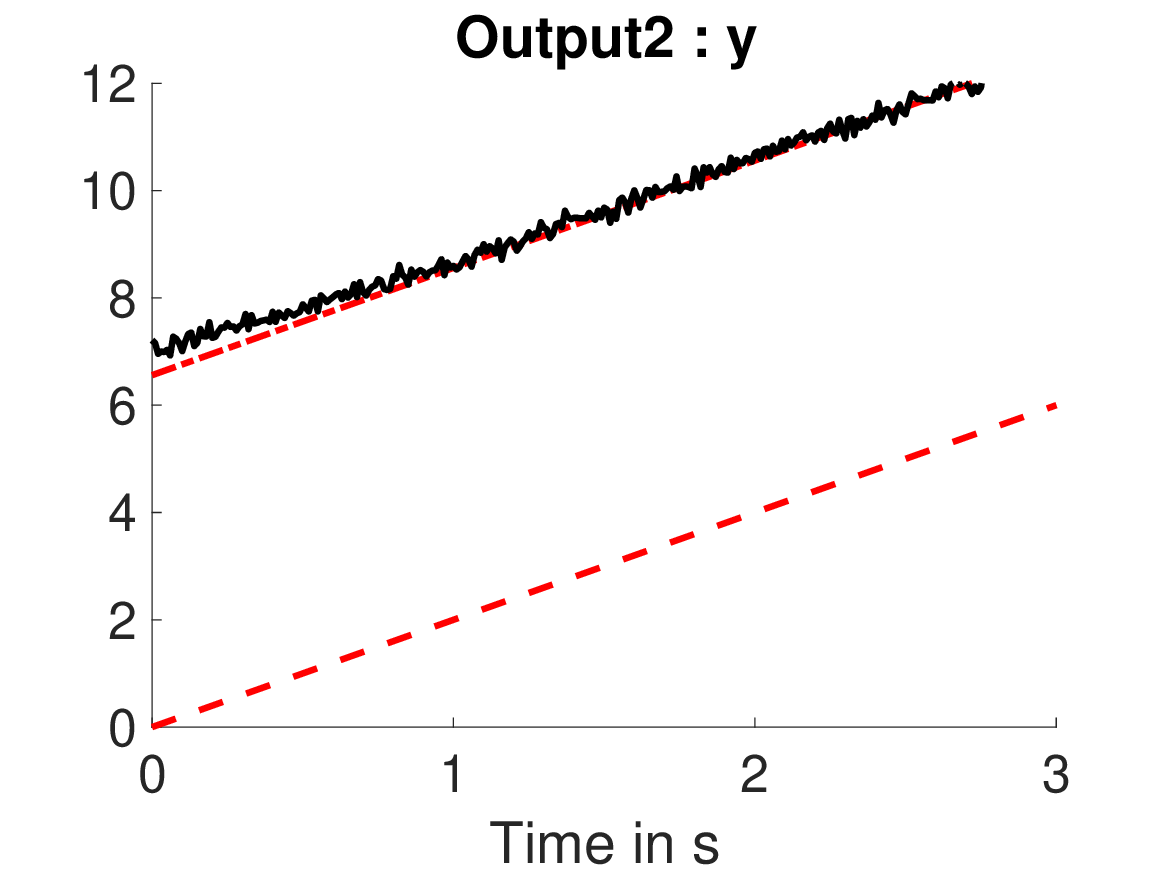,width=0.3\textwidth}}
\subfigure[\footnotesize $(x,y)$]
{\epsfig{figure=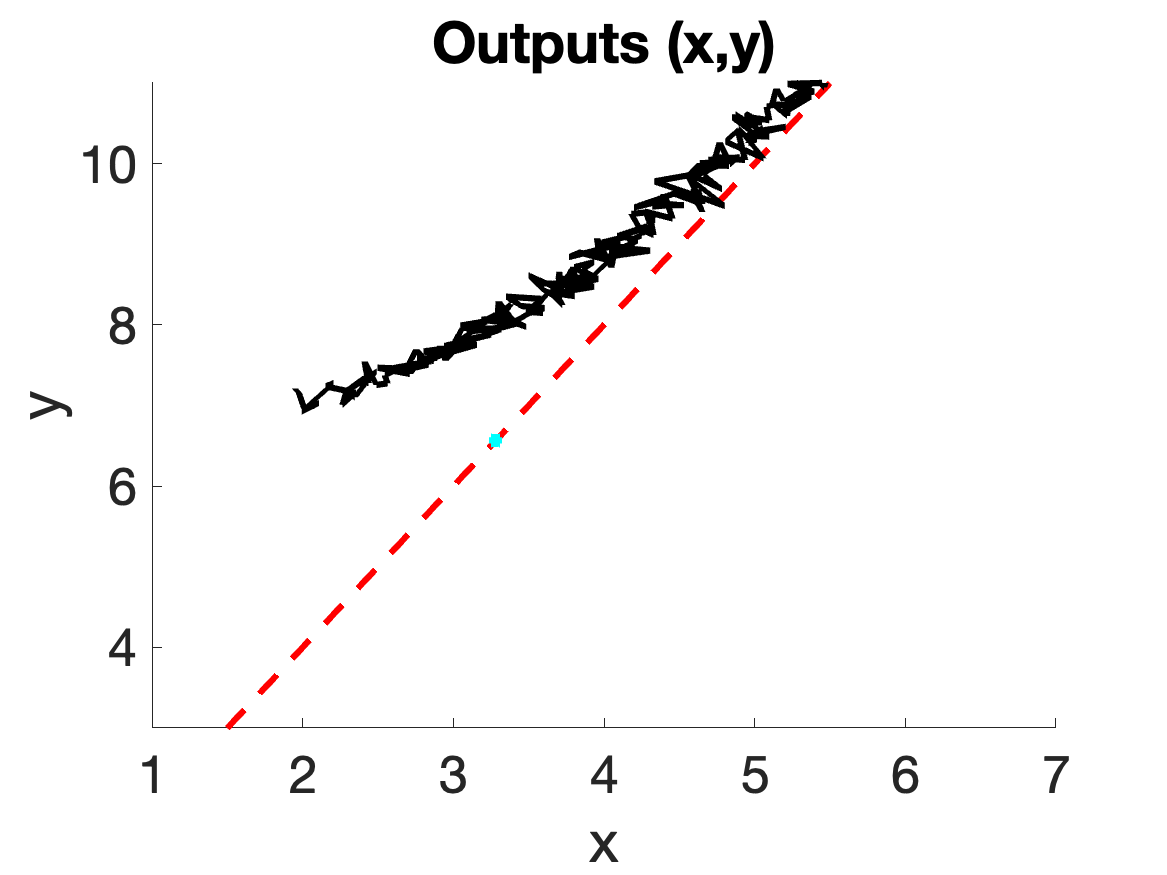,width=0.3\textwidth}}

\caption{Trajectory synchronisation}\label{S}
\end{figure*}

Figures \ref{I} and \ref{S} compare the vehicle's behavior a few moments after it leaves its initial position. In the first case, no synchronization is achieved, and the time lost by reversing in Fig. \ref{I}-(c) is clearly visible. In the second case, the vehicle's trajectory is much smoother (Fig. \ref{I}-(c)). The modified trajectories are plotted in red $\textcolor{red}{-.}$ in Figures \ref{S}-(a) and \ref{S}-(b). The point on the initial trajectory that minimizes distance is plotted in cyan in Fig. \ref{S}-(c).


\subsection{Avoiding unexpected obstacles.}
\begin{figure}[!ht]
{\epsfig{figure=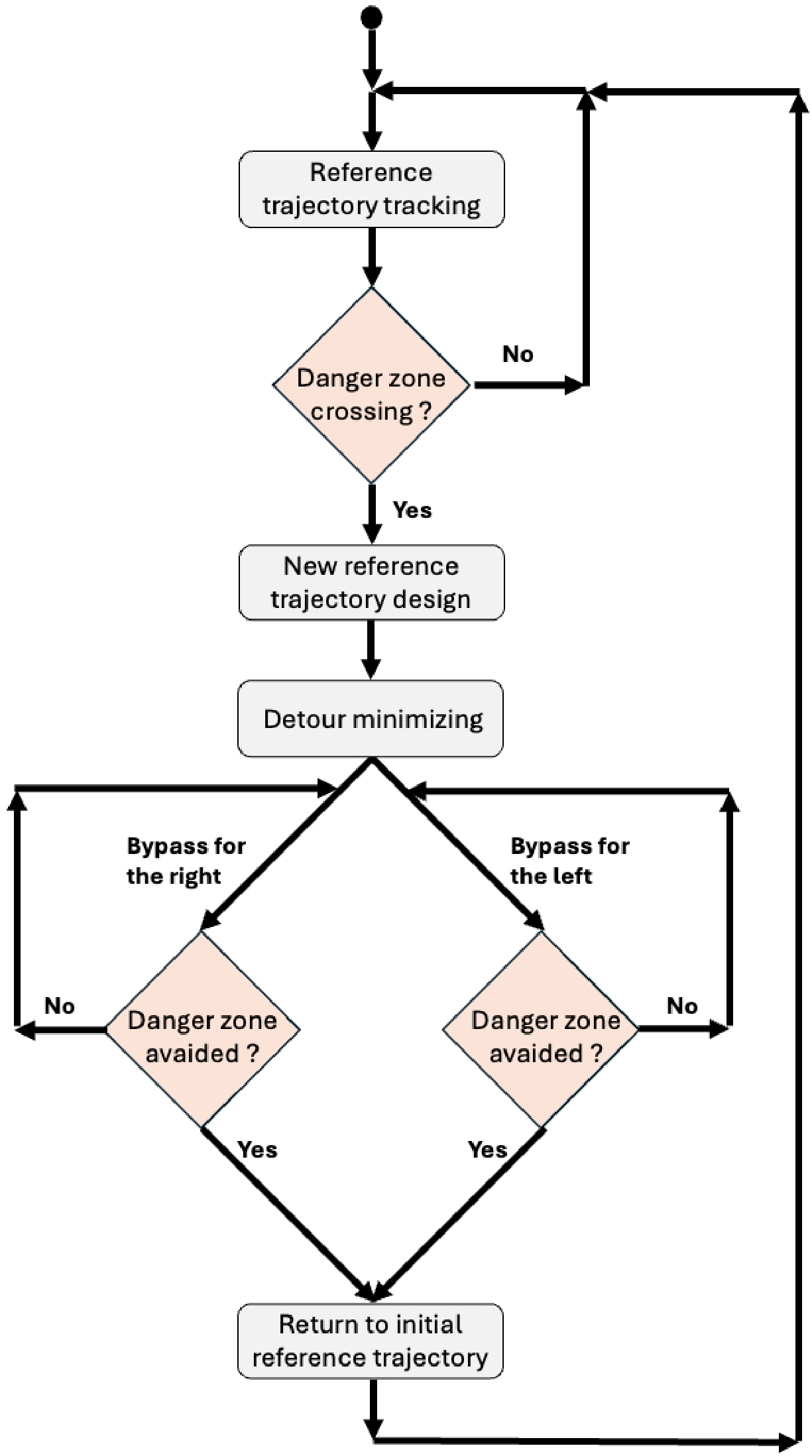,width=0.7\textwidth}}
\centering
\caption{Unexpected obstacle avoidance: Algorithm}\label{algo}
\end{figure}
\begin{figure*}[!ht]
\centering
\subfigure[\footnotesize $u_1$ and $u_{\rm ref,1}$]
{\epsfig{figure=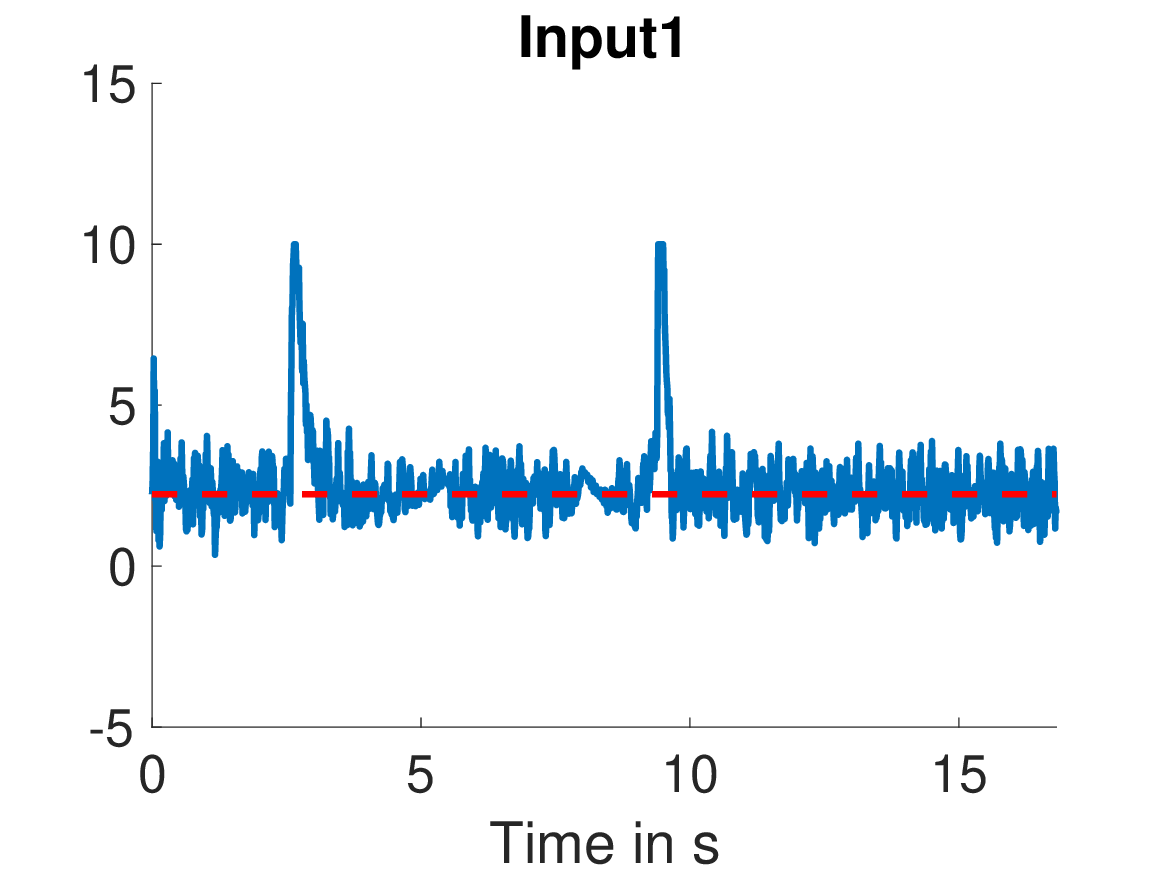,width=0.3\textwidth}}
\subfigure[\footnotesize $u_2$ and $u_{\rm ref,2}$]
{\epsfig{figure=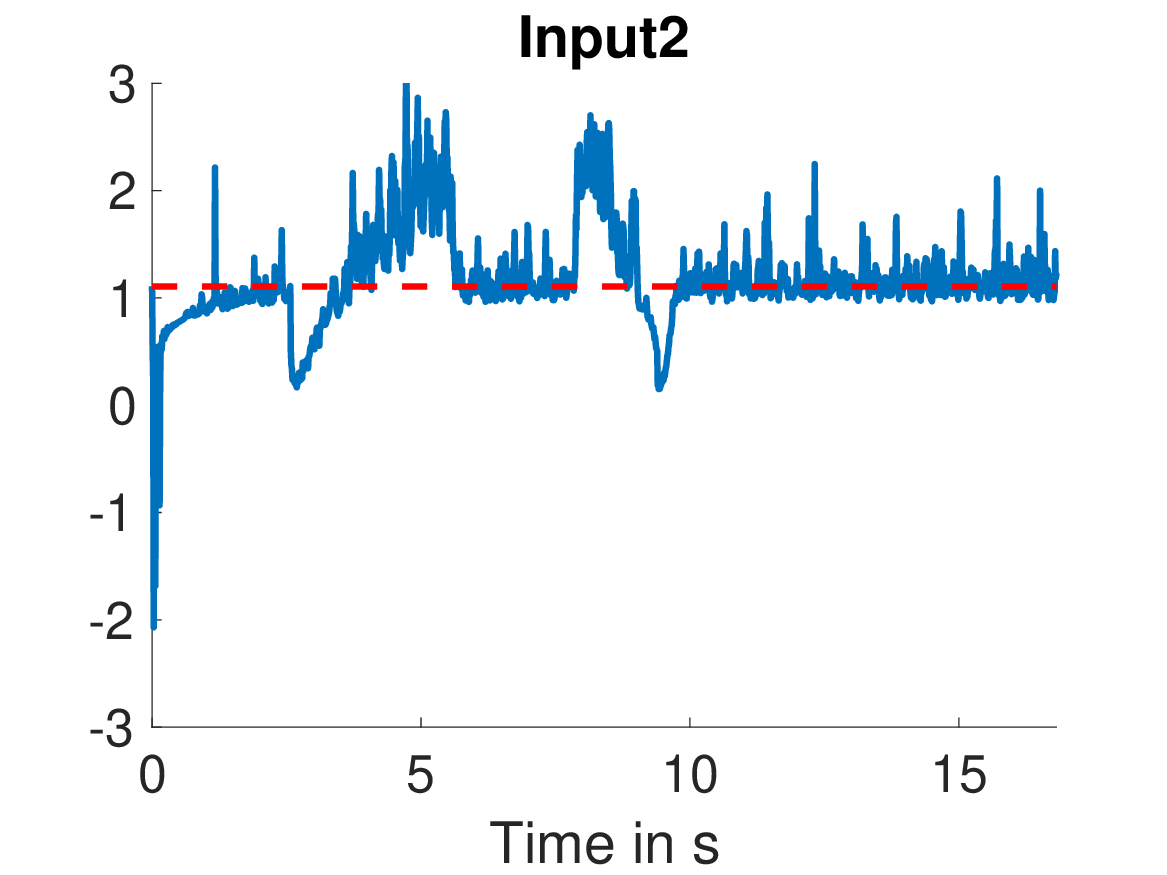,width=0.3\textwidth}}
\caption{HEOL-nominal : inputs}\label{HNU}
\end{figure*}
\begin{figure*}[!ht]
\centering
\subfigure[\footnotesize $x(t)$]
{\epsfig{figure=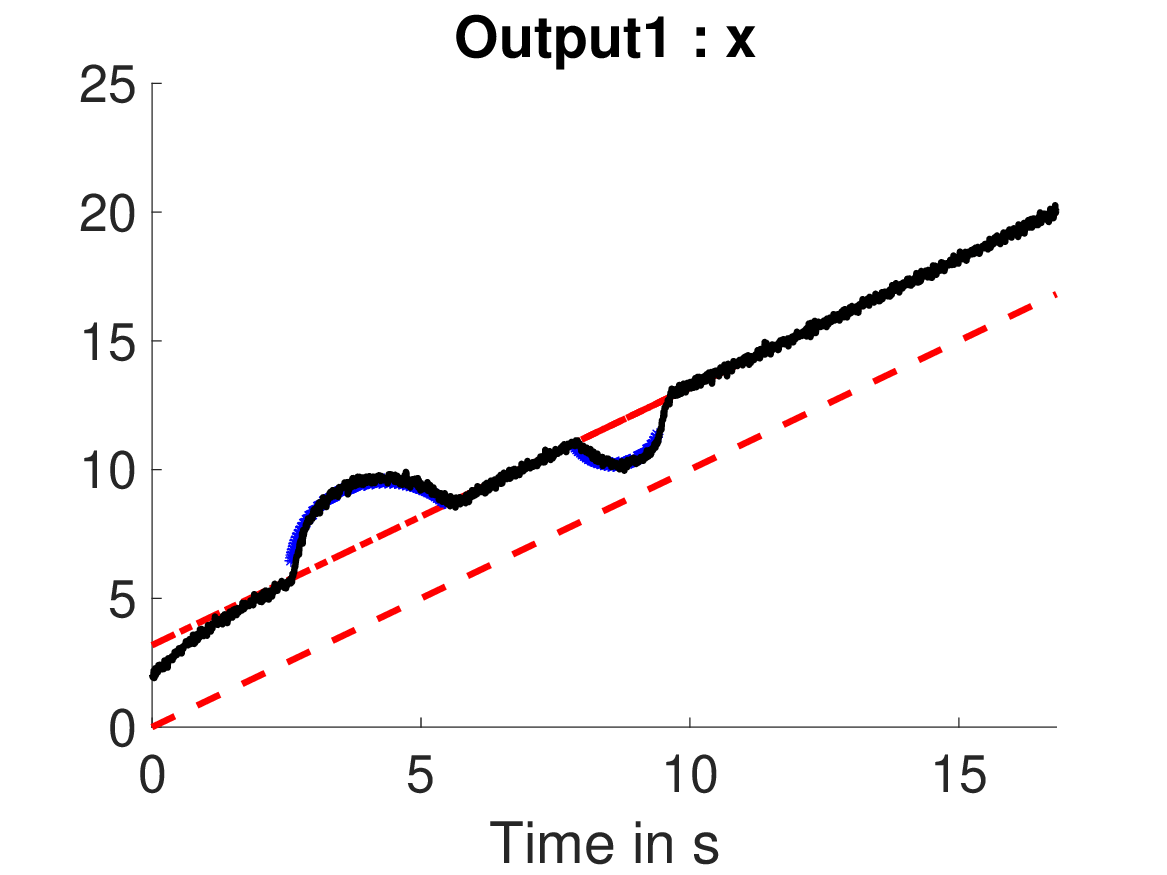,width=0.3\textwidth}}
\subfigure[\footnotesize $y(t)$]
{\epsfig{figure=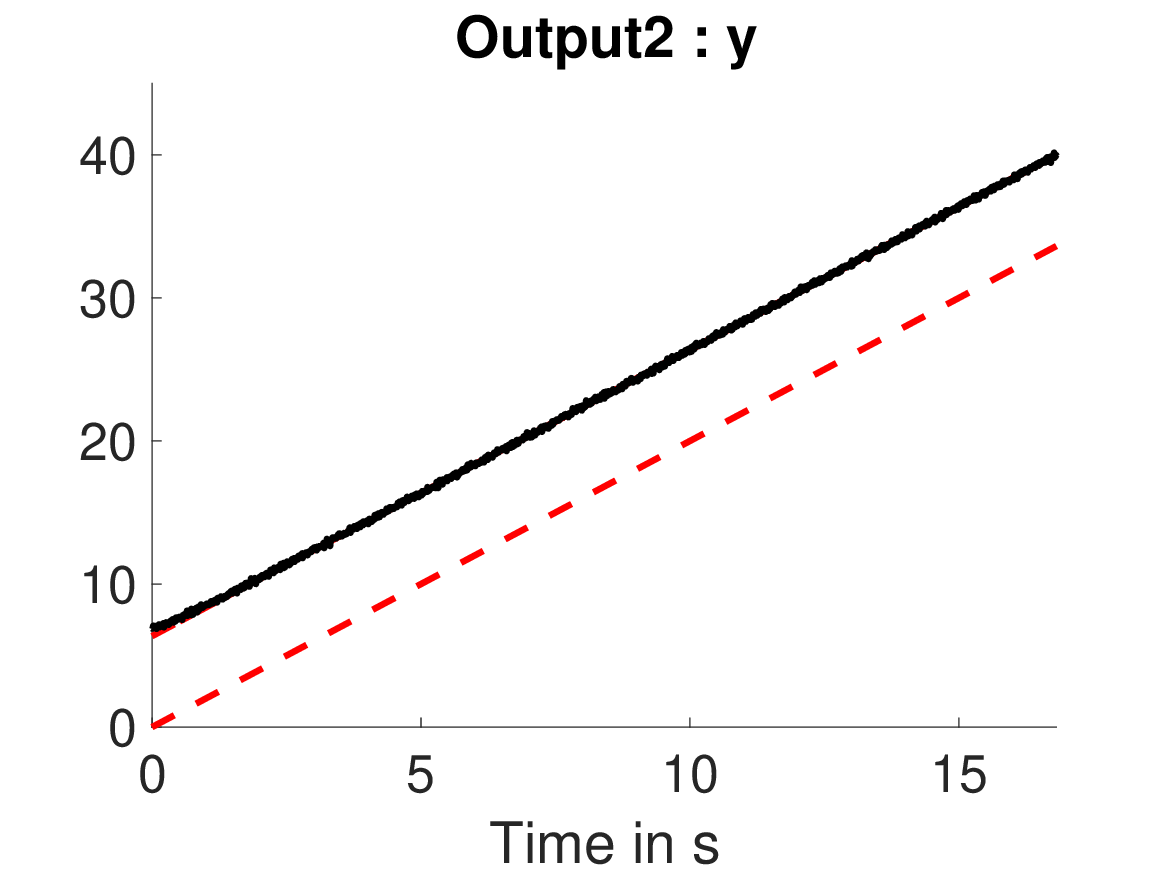,width=0.3\textwidth}}
\subfigure[\footnotesize $(x,y)$]
{\epsfig{figure=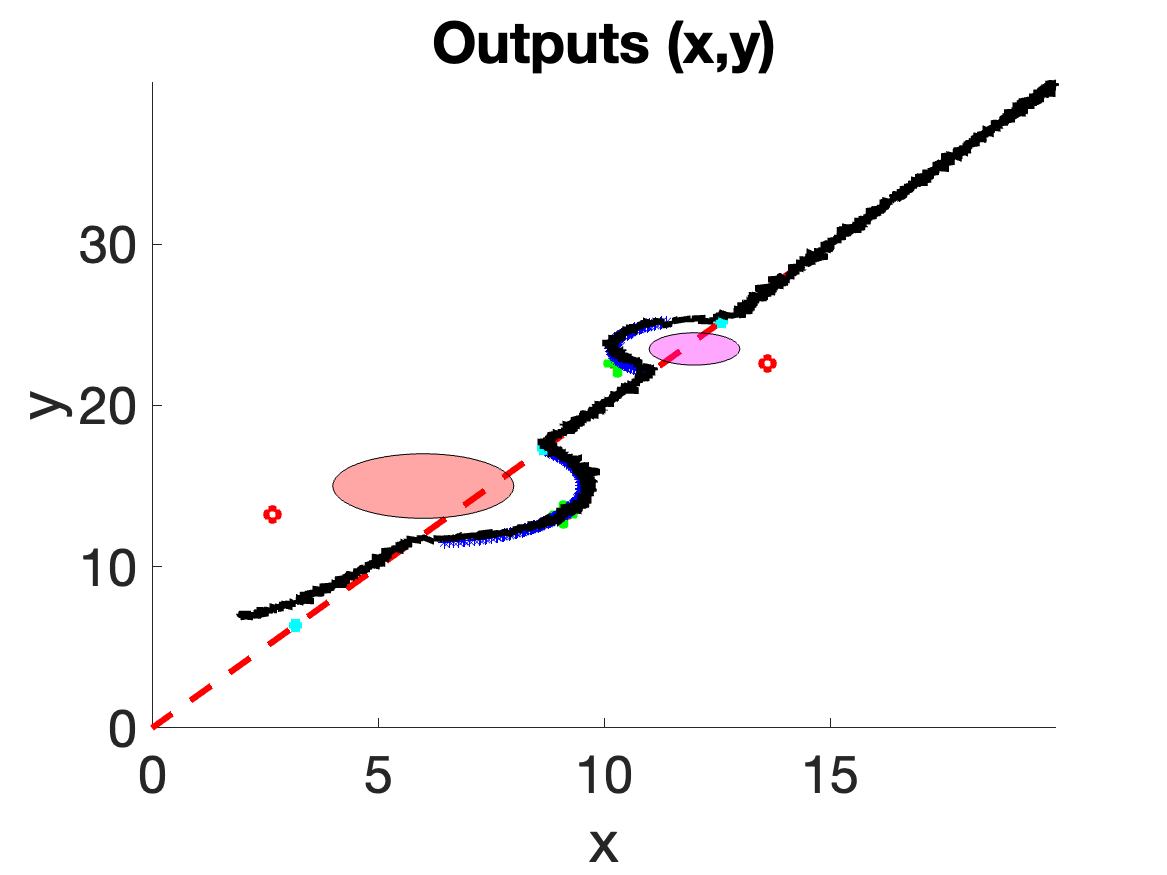,width=0.3\textwidth}}
\caption{HEOL-nominal : Outputs}\label{HNY}
\end{figure*}
\begin{figure*}[!ht]
\centering
\subfigure[\footnotesize $u_1$ ]
{\epsfig{figure=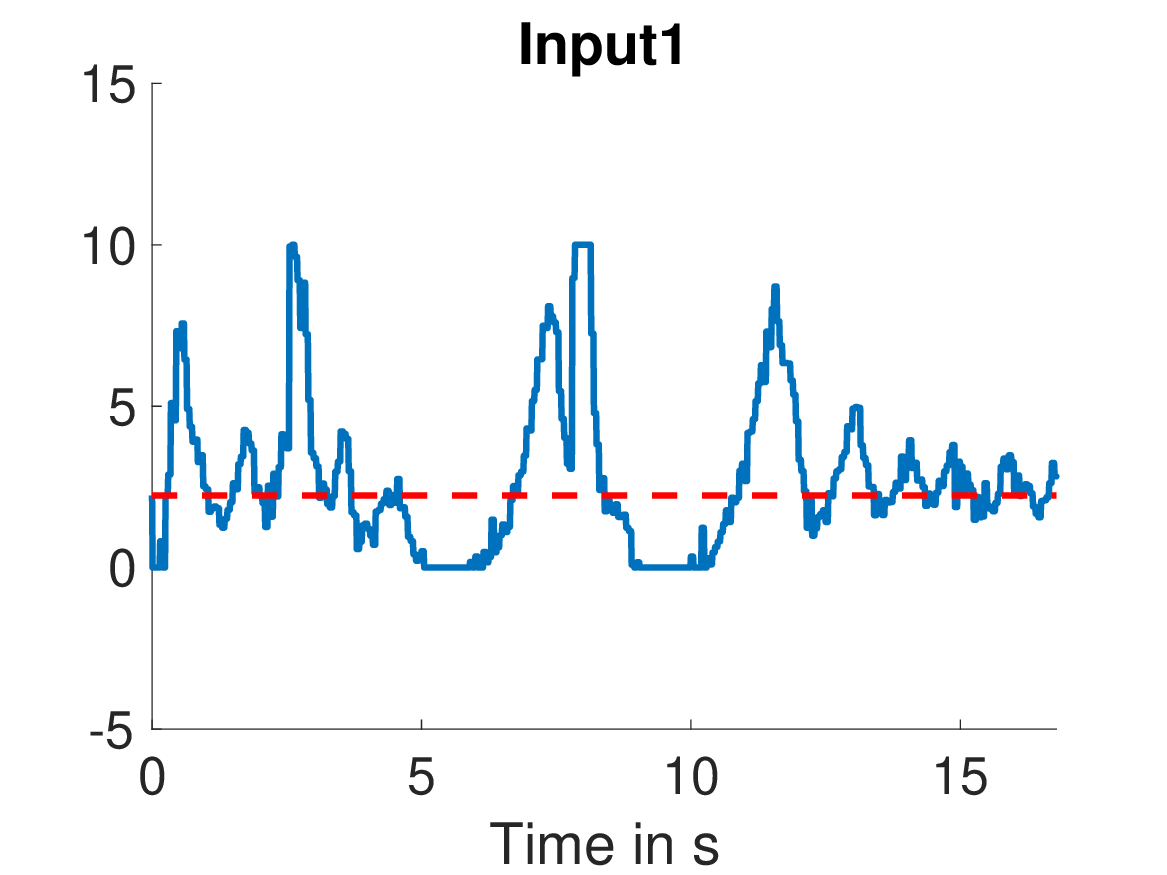,width=0.3\textwidth}}
\subfigure[\footnotesize $u_2$ ]
{\epsfig{figure=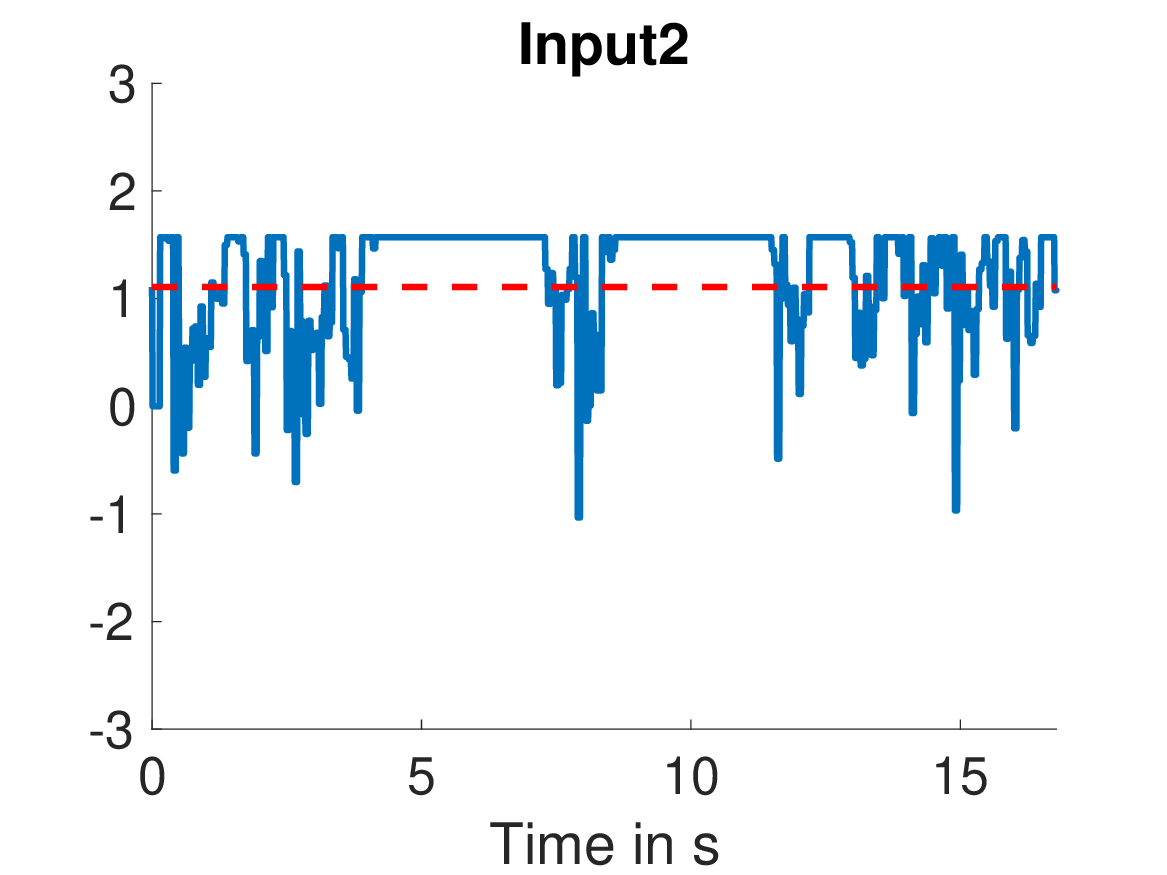,width=0.3\textwidth}}
\caption{MFPC-nominal : inputs}\label{SNU}
\end{figure*}
\begin{figure*}[!ht]
\centering
\subfigure[\footnotesize $x(t)$]
{\epsfig{figure=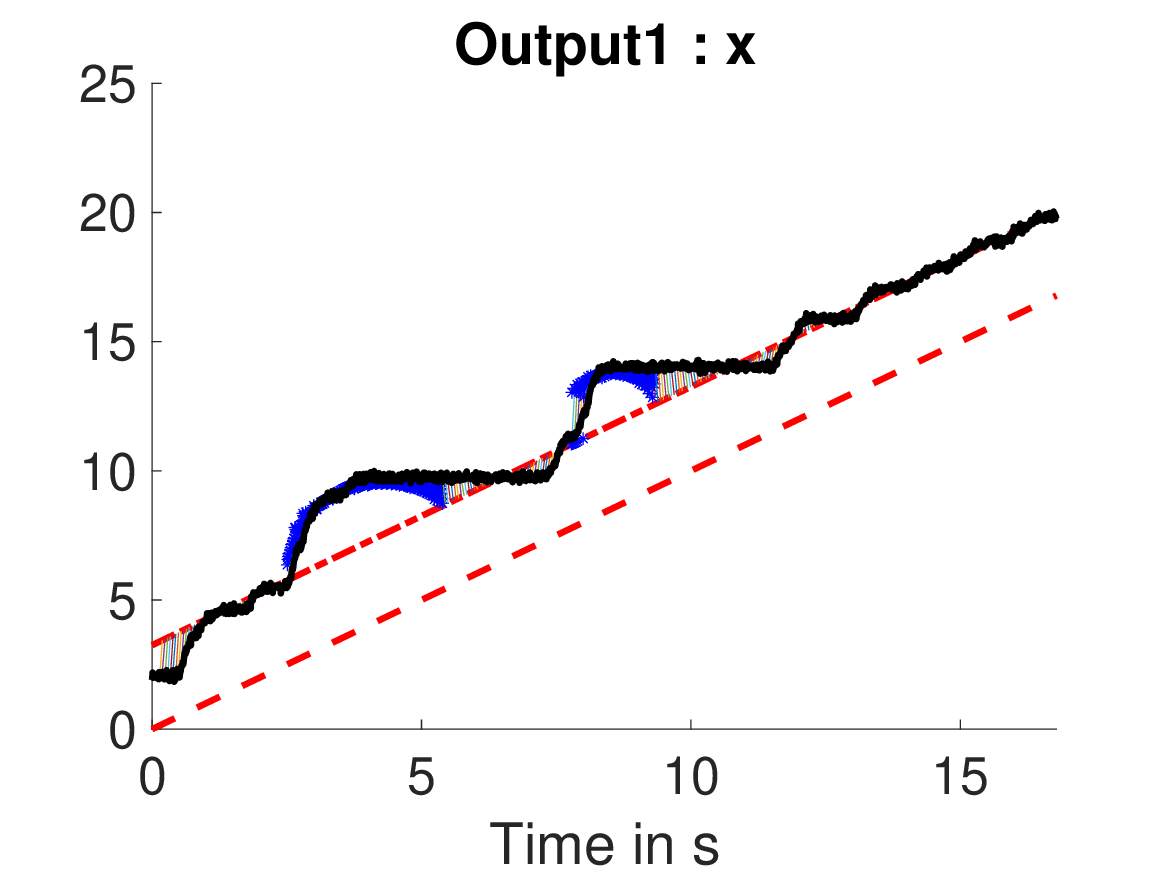,width=0.3\textwidth}}
\subfigure[\footnotesize $y(t)$]
{\epsfig{figure=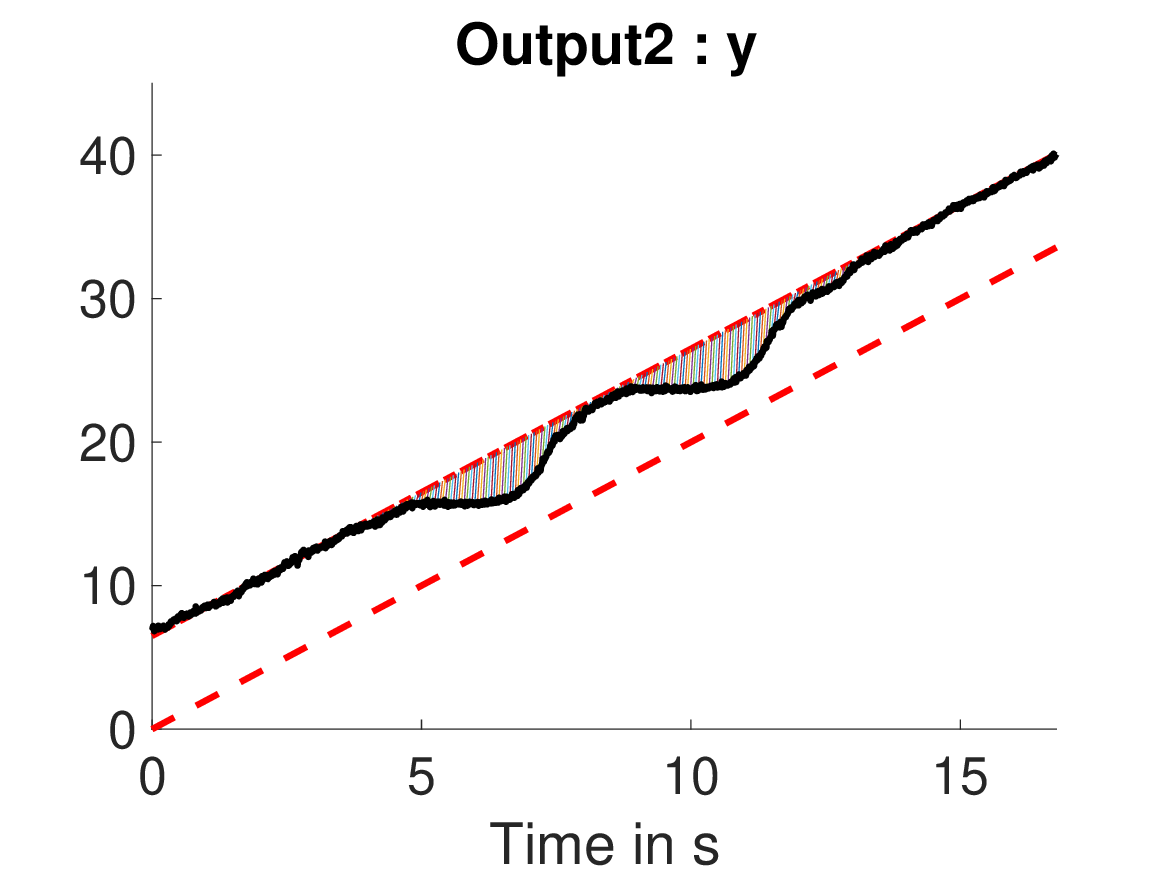,width=0.3\textwidth}}
\subfigure[\footnotesize $(x,y)$]
{\epsfig{figure=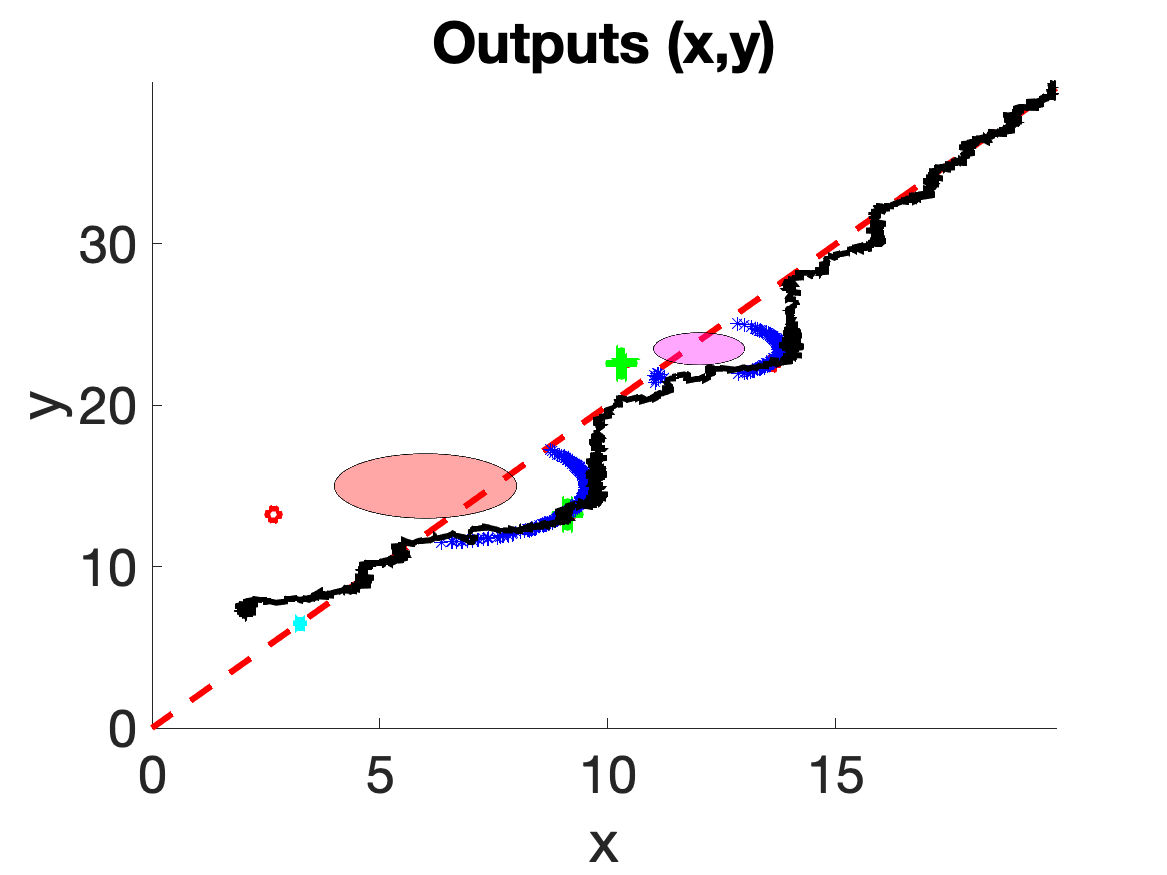,width=0.3\textwidth}}
\caption{MFPC-nominal : Outputs}\label{SNY}
\end{figure*}
\begin{figure*}[!ht]
\centering
\subfigure[\footnotesize $x(t)$]
{\epsfig{figure=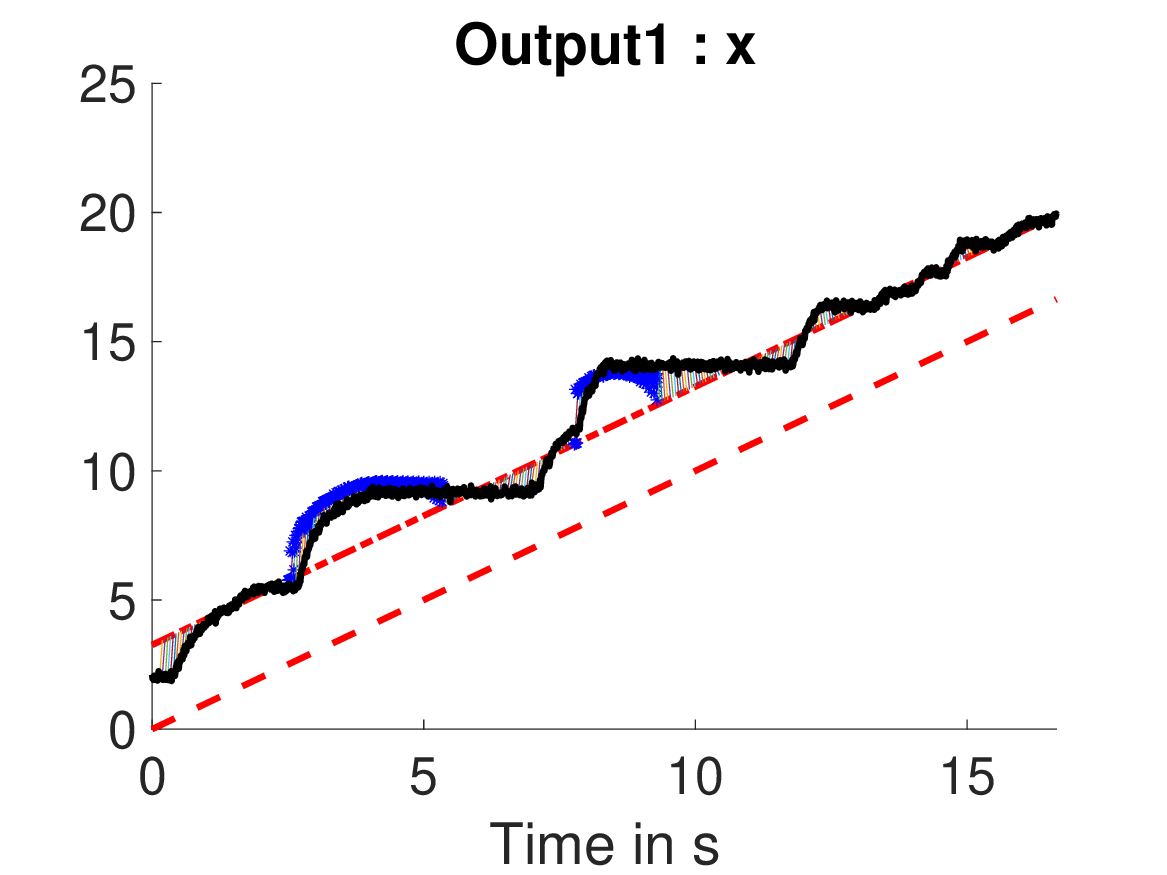,width=0.3\textwidth}}
\subfigure[\footnotesize $y(t)$]
{\epsfig{figure=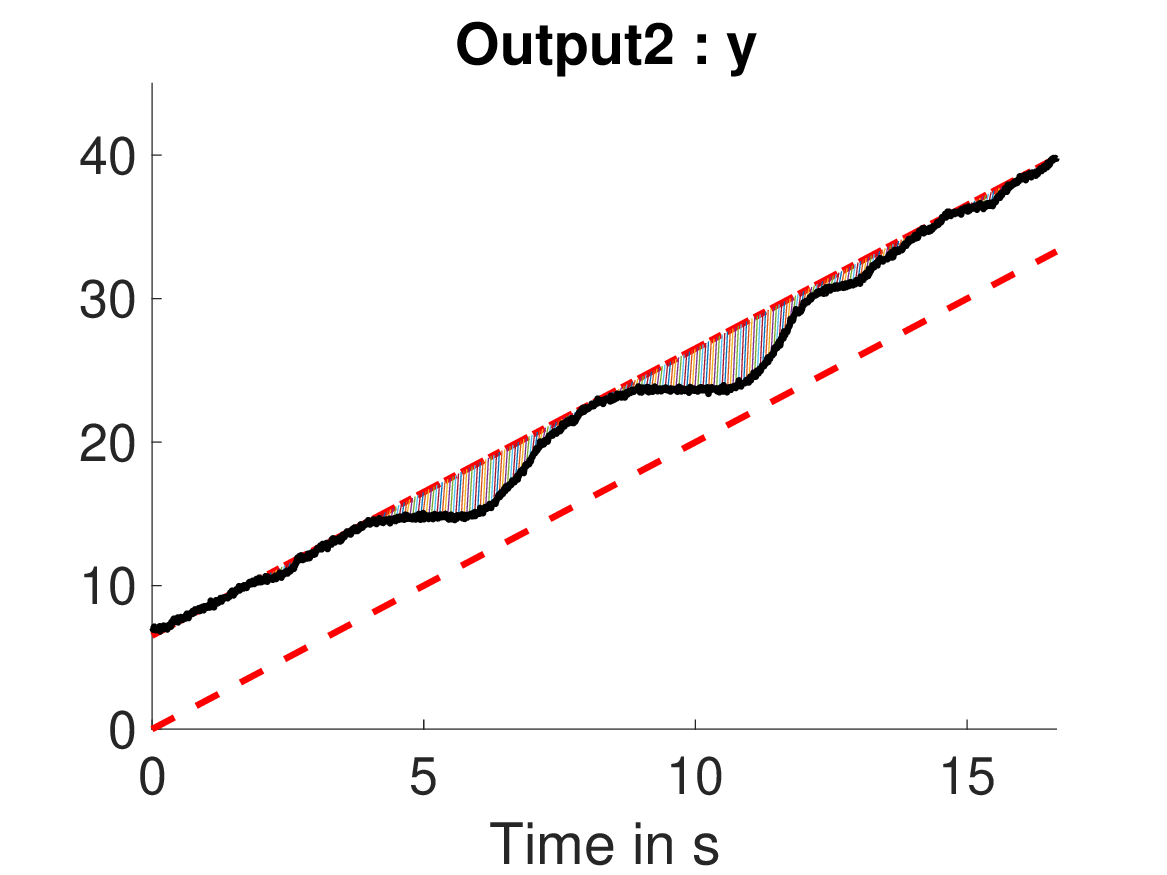,width=0.3\textwidth}}
\subfigure[\footnotesize $(x,y)$]
{\epsfig{figure=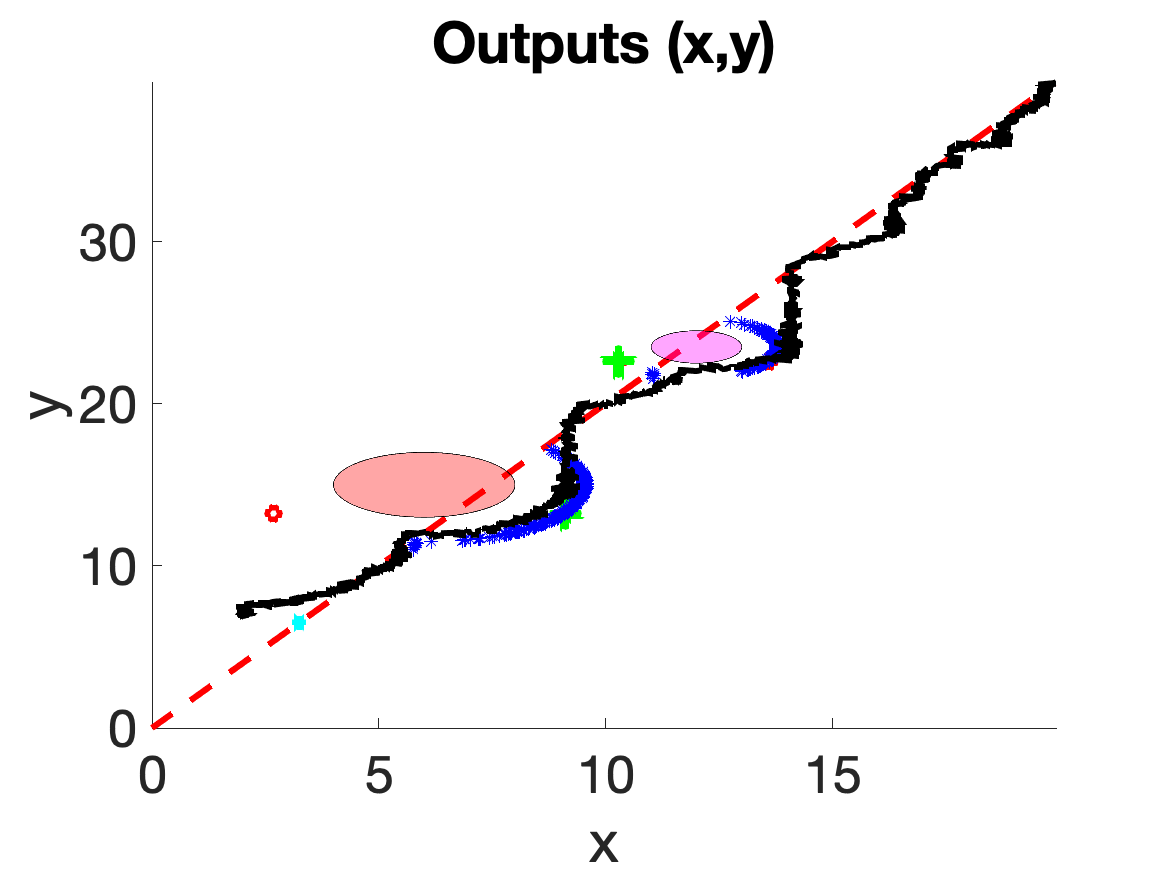,width=0.3\textwidth}}
\caption{MFPC-perturbed : Outputs}\label{SP2Y}
\end{figure*}

The algorithm outline displayed by Figure \ref{algo} shows the central         steps involved in avoiding unexpected obstacles. If there are no obstacles, the vehicle follows the initially defined reference trajectory. However, if the reference path crosses a danger zone (defined by the obstacle's contours plus a safety margin), the path must be modified. The obstacle can be bypassed to the right or left; the optimal choice is made by minimizing the detour. Once the obstacle has been bypassed, the vehicle continues on the initial trajectory. The results obtained are illustrated in Figures \ref{HNU}-\ref{SP2Y}. The avoidance contours added with a safety margin, are plotted in blue: see Fig. \ref{HNY}-(a) - \ref{HNY}-(c) for HEOL and Fig. \ref{SNY}-(a) - \ref{SNY}-(c) for MFPC. They display excellent behavior, even in noisy environments ($\mathcal{N}(0,0.1)$), and they are very adaptable, adjusting to new situations quickly.

In the MFPC case, bypasses are always performed on the right side to satisfy the $u_2$ constraint. Although this requires a larger detour, it is still satisfactory. Furthermore, as demonstrated in Figures \ref{SNY}-(c) and \ref{SP2Y}-(c), the vehicle passes relatively close to the obstacle. This could be improved by increasing the safety margin.

\subsection{Robustness}
\begin{figure*}[!ht]
\centering
\subfigure[\footnotesize $u_1$ and $u_{\rm ref,1}$]
{\epsfig{figure=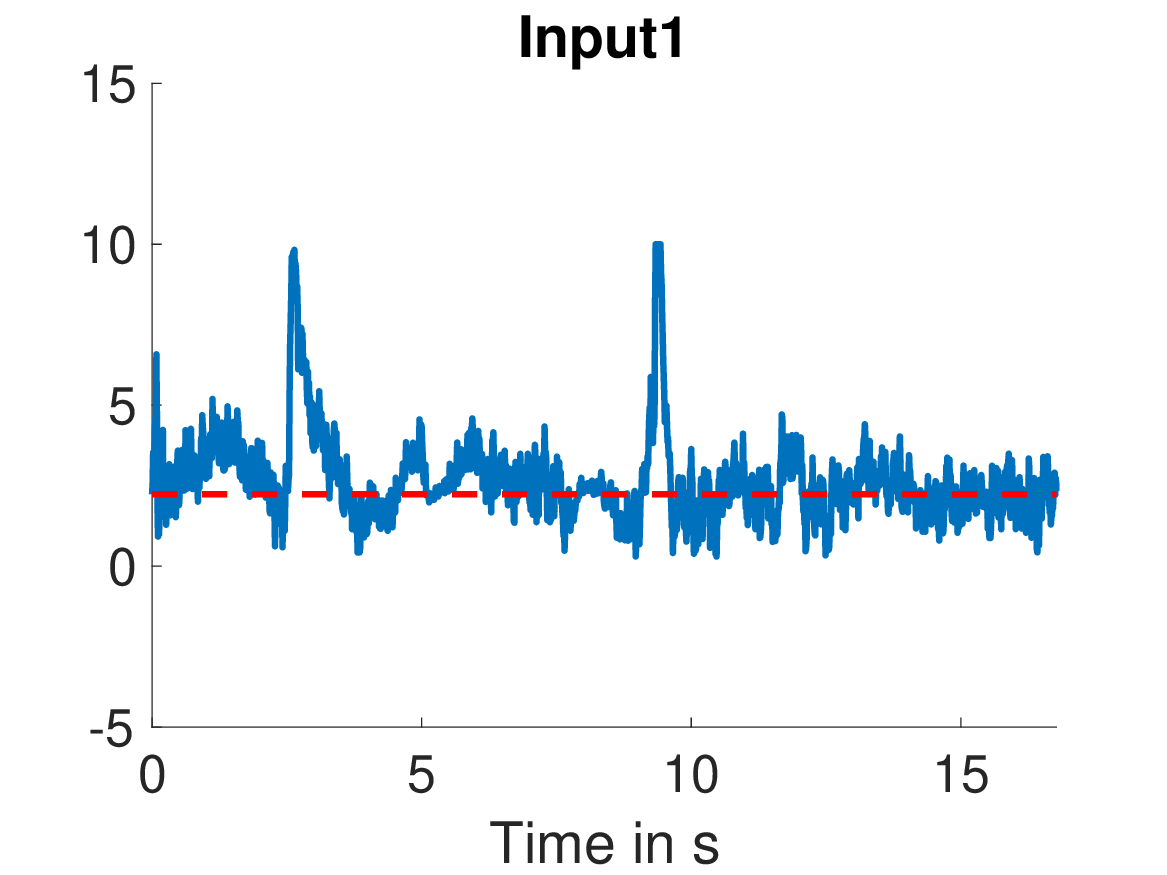,width=0.3\textwidth}}
\subfigure[\footnotesize $u_2$ and $u_{\rm ref,2}$]
{\epsfig{figure=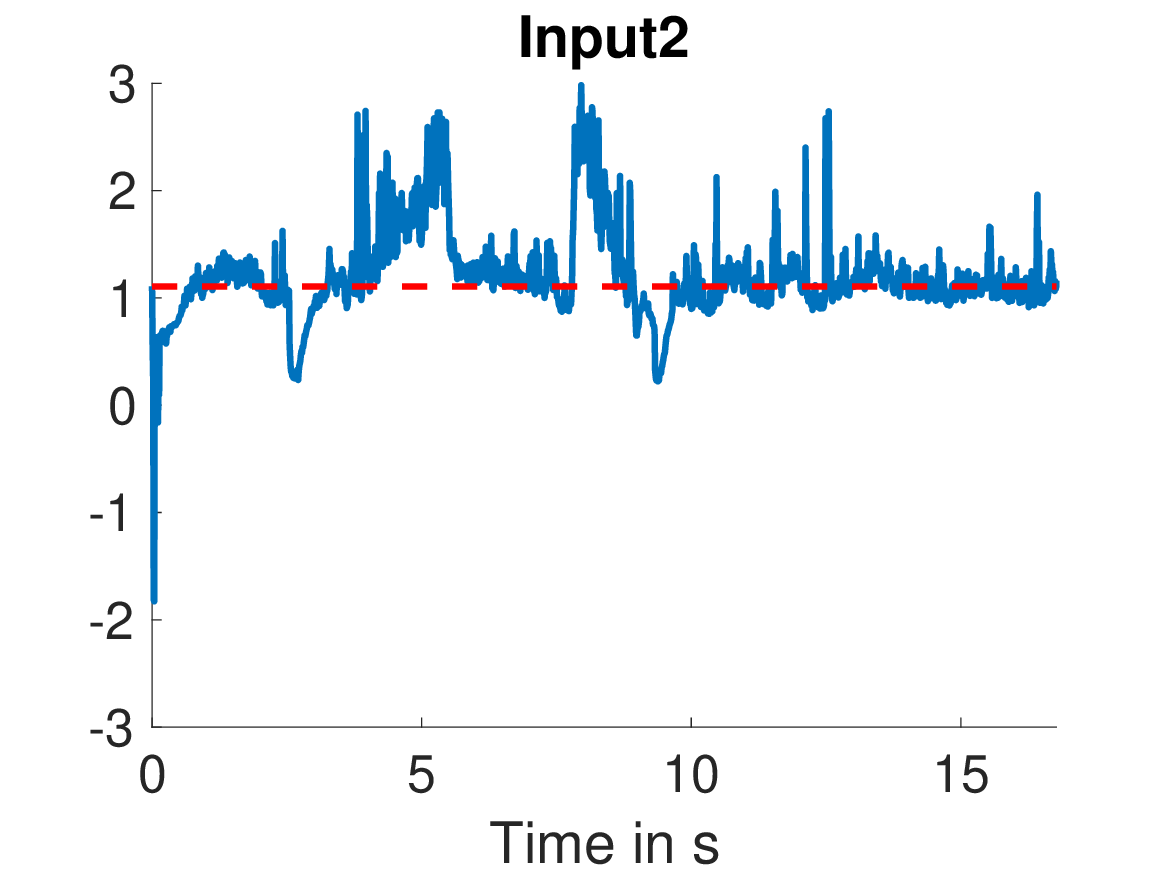,width=0.3\textwidth}}
\subfigure[\footnotesize $p$]
{\epsfig{figure=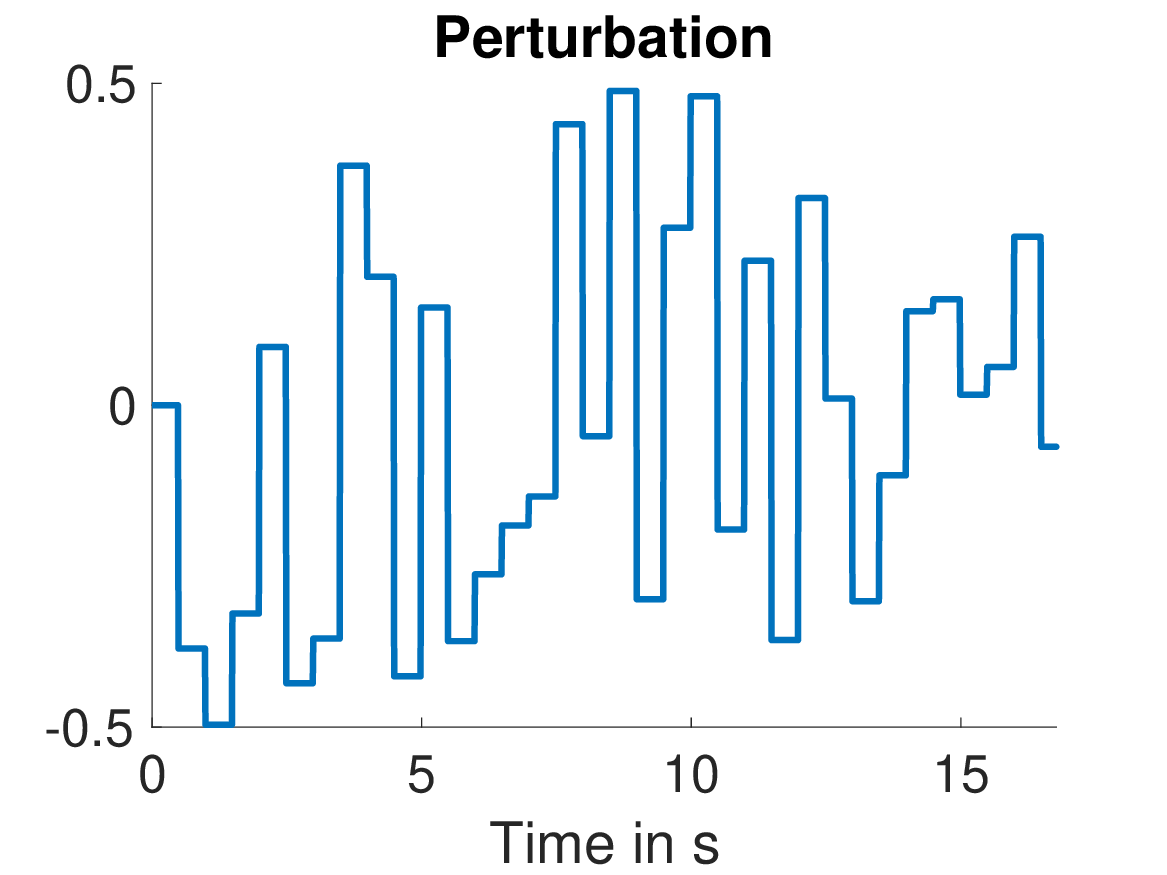,width=0.3\textwidth}}
\caption{HEOL-perturbed : inputs}\label{HP2U}
\end{figure*}

\begin{figure*}[!ht]
\centering
\subfigure[\footnotesize $x(t)$]
{\epsfig{figure=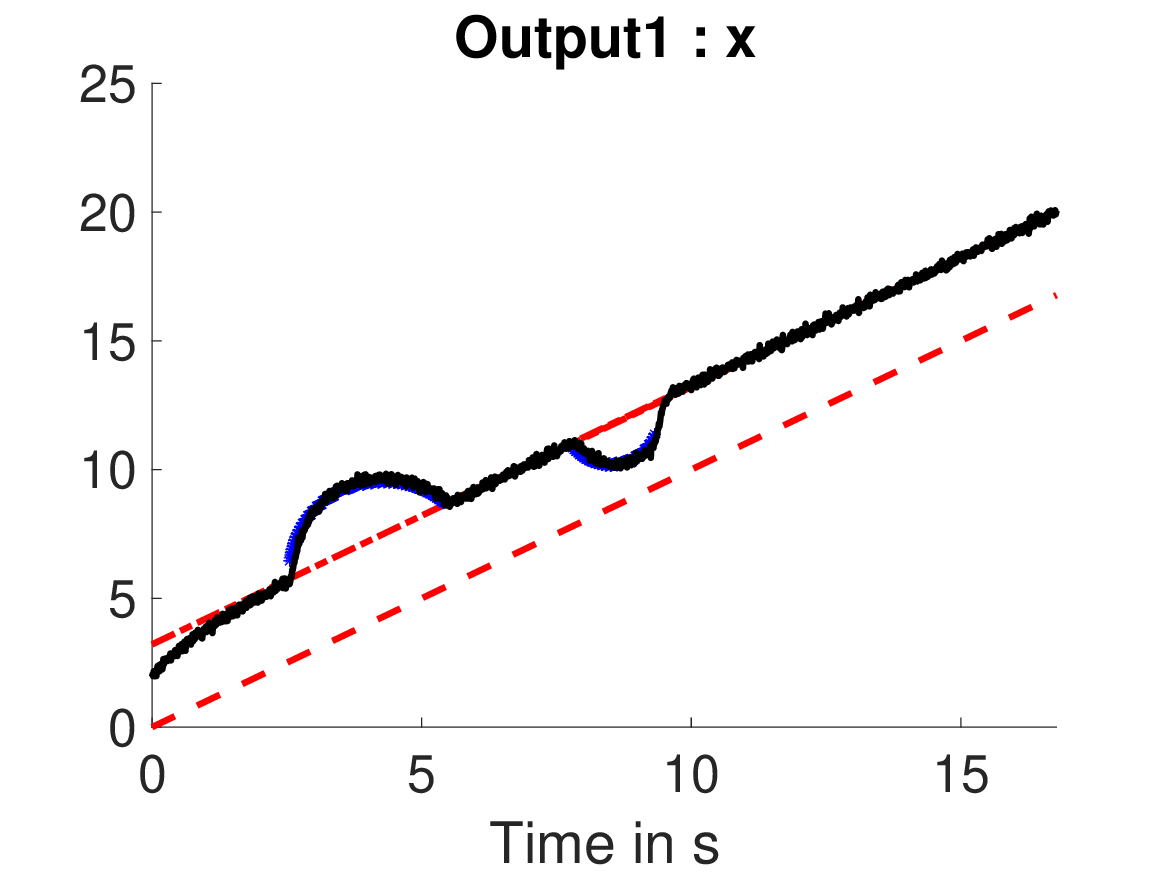,width=0.3\textwidth}}
\subfigure[\footnotesize $y(t)$]
{\epsfig{figure=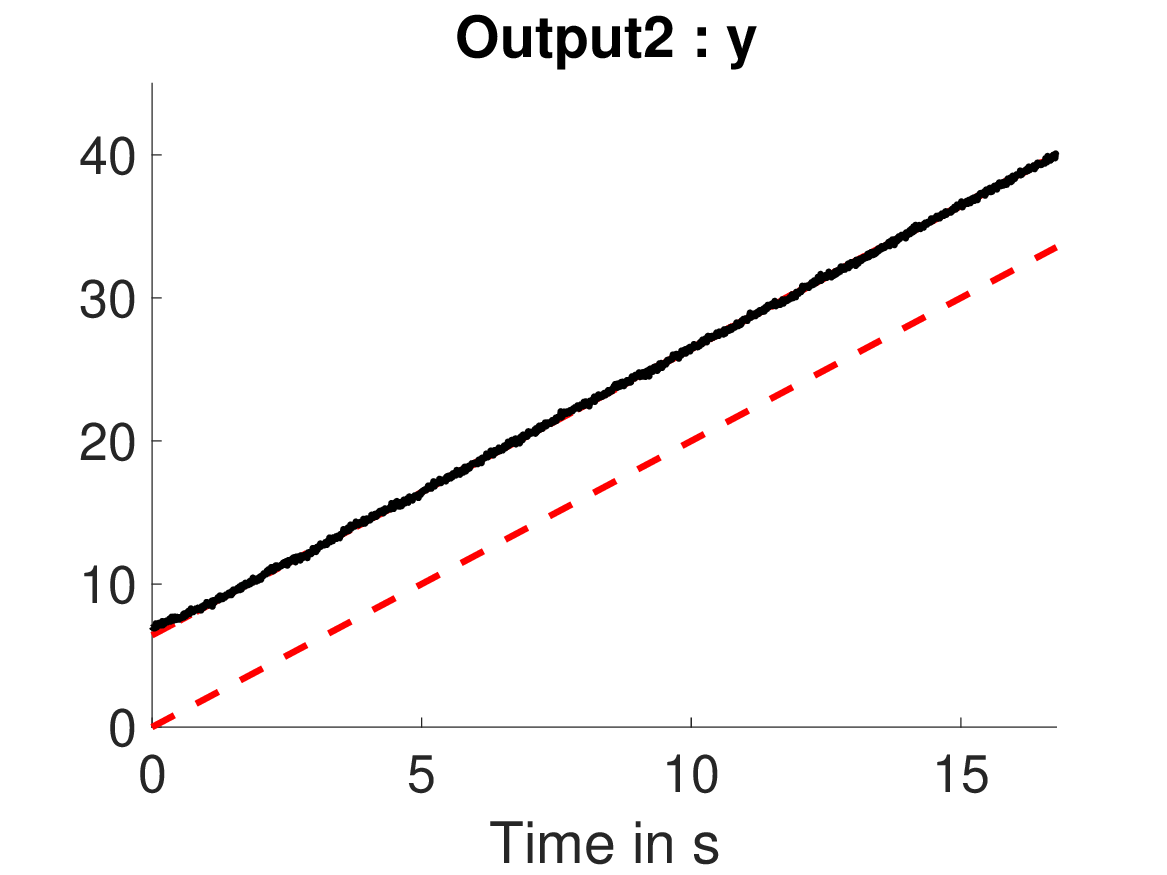,width=0.3\textwidth}}
\subfigure[\footnotesize $(x,y)$]
{\epsfig{figure=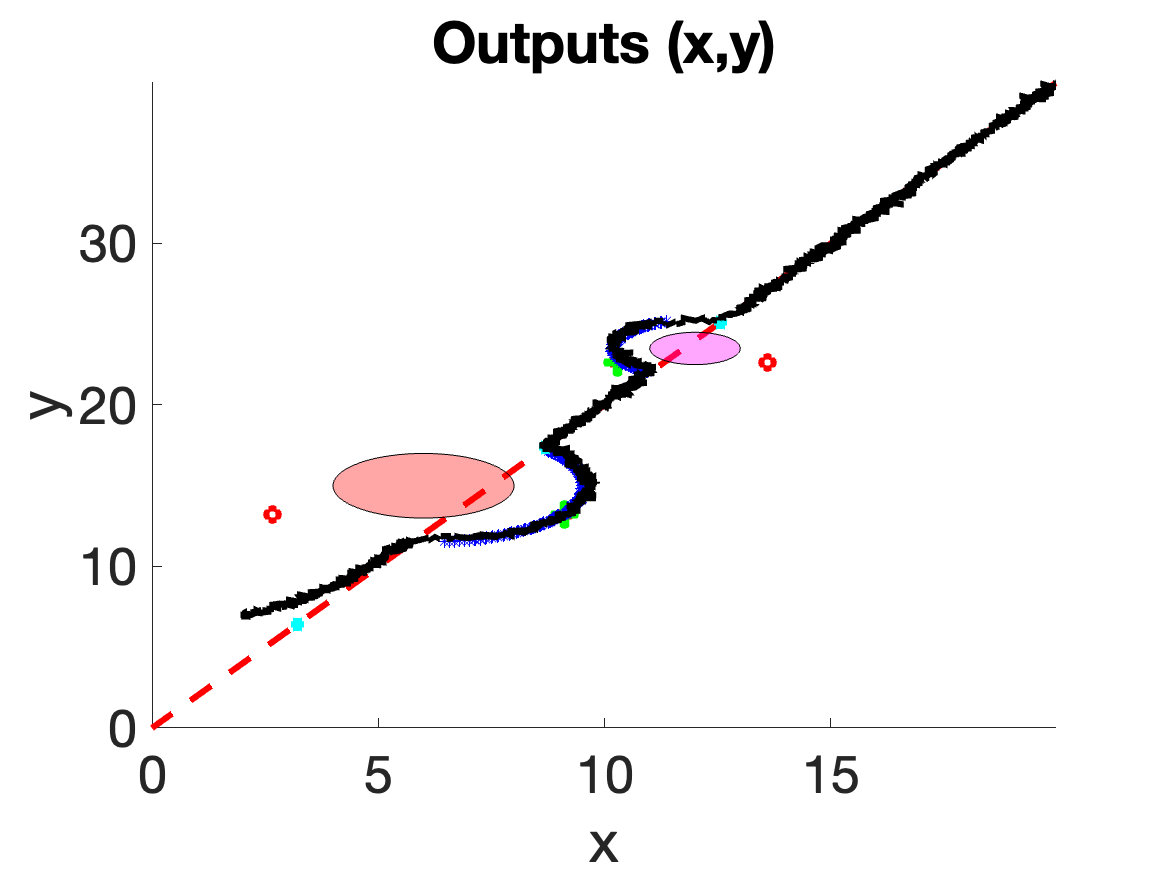,width=0.3\textwidth}}
\caption{HEOL-perturbed : Outputs}\label{HP2Y}
\end{figure*}

\begin{figure*}[!ht]
\centering
\subfigure[\footnotesize $u_1$ ]
{\epsfig{figure=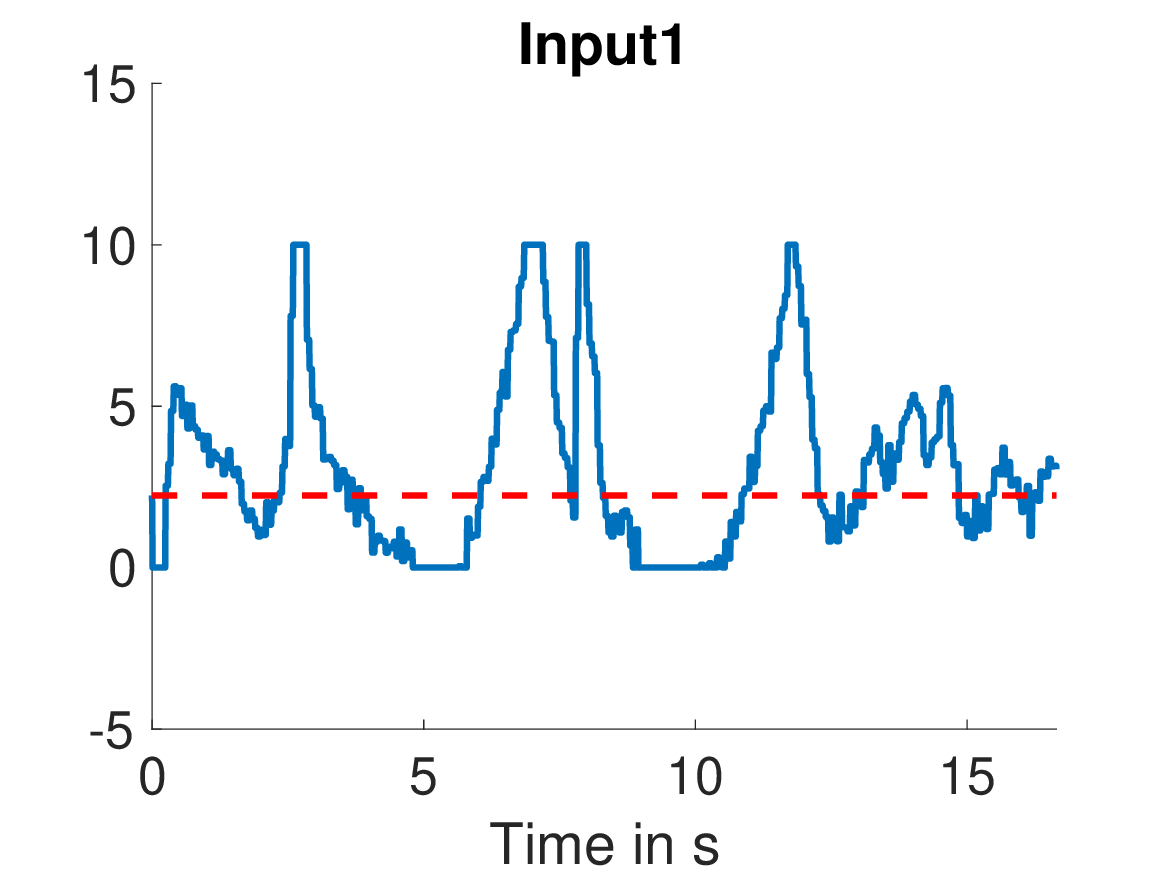,width=0.3\textwidth}}
\subfigure[\footnotesize $u_2$ ]
{\epsfig{figure=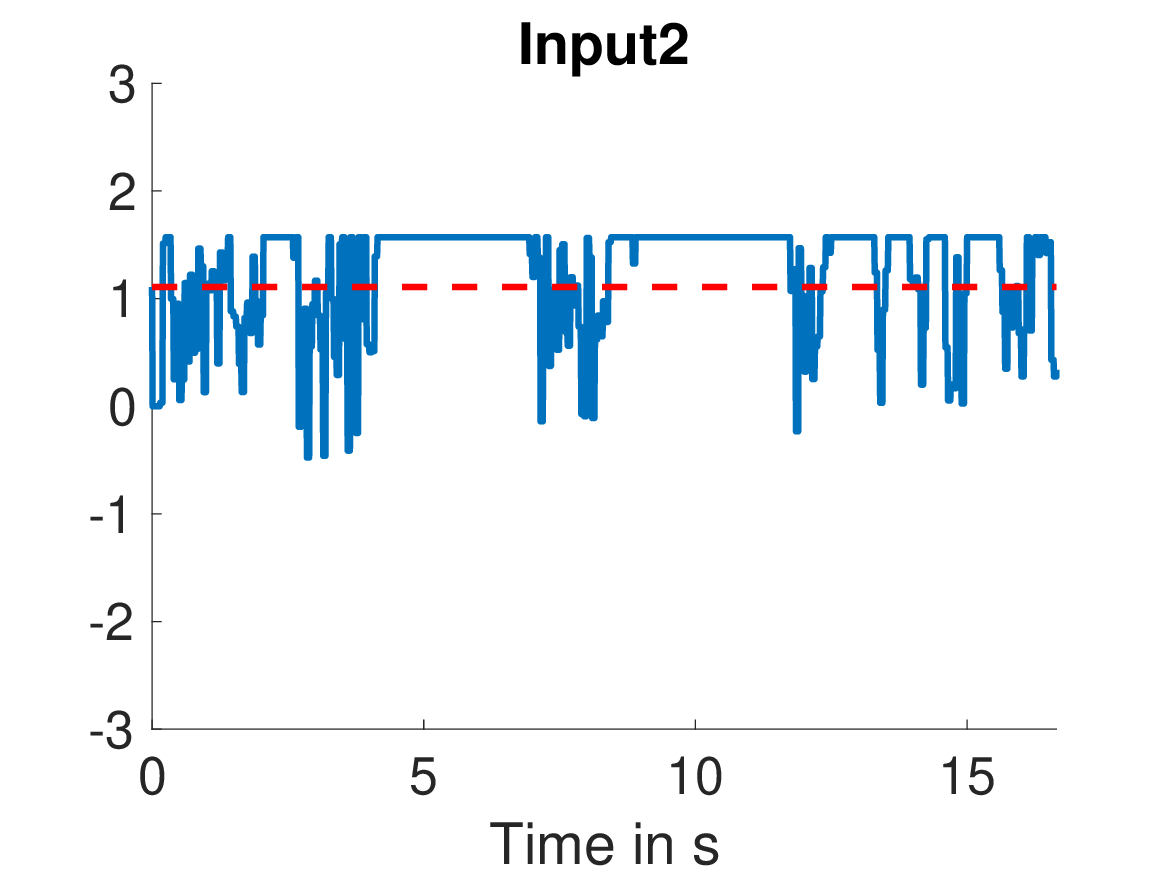,width=0.3\textwidth}}
\subfigure[\footnotesize $p$]
{\epsfig{figure=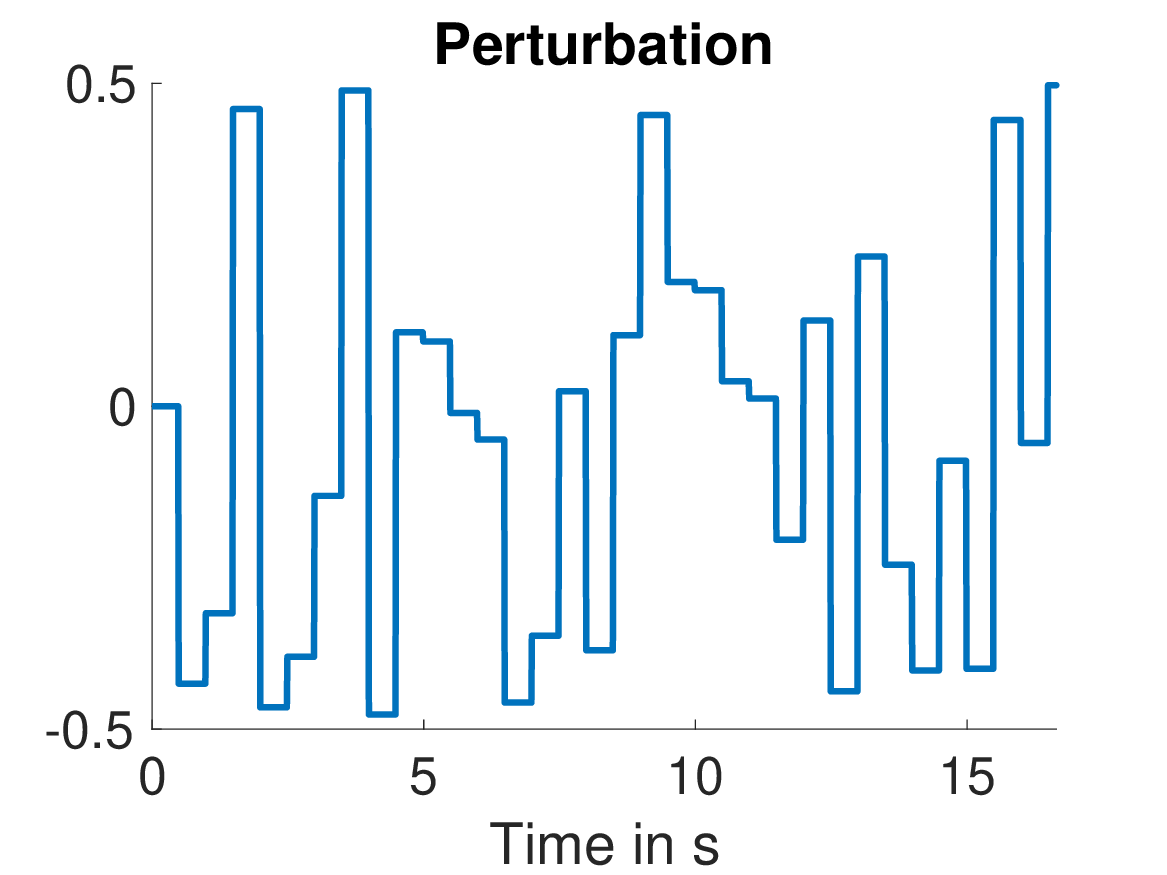,width=0.3\textwidth}}
\caption{MFPC-perturbed : inputs}\label{SP2U}
\end{figure*}

To assess the robustness of the proposed control scheme, an unknown external disturbance is introduced to the dynamics of $y$. This disturbance could correspond to the effect of a crosswind, for example. Thus, the vehicle behavior model becomes
\begin{equation}
\begin{cases}
    \dot x=u_1\cos(u_2)\\
    \dot y=u_1(1+p)\sin(u_2)
\end{cases}
\end{equation}
The perturbation $p$ is a piecewise constant function and its values are taken from a uniform distribution on $[-0.5, +0.5]$.
In the HEOL case, where the model is used, the relationships between the auxiliary control variables and the true control variables are no longer verified. As Fig. \ref{HP2U} and \ref{HP2Y} show, trajectory tracking still performs very well.
In the MFPC case, where no model knowledge is used, the commands are slightly more dynamic, though the tracking quality remains unchanged (see Fig. \ref{SP2U} and \ref{SP2Y}).
It should be noted that this perturbation is unpredictable and non-reproducible, which complicates its integration into conventional learning methods.

\section{Conclusion}\label{conclu}
The above promising results require, of course, confirmation and expansion in subsequent studies. One example is stability. In the case of the combination of flatness-based control and HEOL, local stability is straightforward. However, nothing has been demonstrated yet for MFPC.

The low computing burden required by our methods could be particularly useful for robotics of more complex systems (see, e.g., \cite{uav}) and/or of deformable bodies and, more generally, bodies that are difficult to model (see, e.g., \cite{berenson,yin}). 



\end{document}